%% file: fortier_article_version.tex
\documentclass[structabstract]{aa}  
\usepackage{graphicx,natbib}
\usepackage{ulem}
\usepackage[latin1]{inputenc}
\usepackage[T1]{fontenc}
\bibpunct{(}{)}{;}{a}{}{,}
\usepackage{txfonts}
\include{def}

\usepackage{changebar}
\begin{document}
   \title{Planet formation models: the interplay with the planetesimal disc}

   
   \author{A. Fortier \inst{1}  
       \and
       Y. Alibert \inst{1, 2}
       \and
             F. Carron \inst{1}
        \and
       W. Benz \inst{1}
       \and
        	K-M. Dittkrist\inst{3}
     }
\offprints{A. Fortier \\ \email andrea.fortier@space.unibe.ch}
\institute{Center for Space and Habitability \& Physikalisches Institut, Universitaet Bern, CH-3012 Bern, Switzerland
\and
Observatoire de Besan\c con, 41 avenue de l'Observatoire, 25000 Besan\c con, France
\and
Max-Planck Institute for Astronomy, K\" onigstuhl 17, 69117 Heidelberg, Germany}

         %


   \date{;}

 
  \abstract
   {According to the sequential accretion model (or core nucleated accretion model), giant planet formation is based first on the formation of a solid core which, when massive enough,  can gravitationally bind gas from the nebula to form the envelope. The most critical part of the model is the formation time of the core: in order to trigger the accretion of gas the core has to grow up to several Earth masses before the gas component of the protoplanetary disc dissipates. }
    {In this paper, we calculate planetary formation models including a detailed description of the dynamics of the planetesimal disc, taking into account both gas drag and excitation of forming planets.}
   {We compute the formation of planets, considering the oligarchic regime for the growth of the solid core.  Embryos growing in the disc stir their neighbour planetesimals, exciting their relative velocities, which makes accretion more difficult. Here we introduce a more realistic  treatment for the evolution of planetesimals' relative velocities, which directly impact on the formation timescale.  For this, we compute the excitation state of planetesimals, as a result of stirring by forming planets, and gas-solid interactions}
   {We find that the formation of giant planets is favoured by the accretion of small planetesimals, as their random velocities are more easily damped by the gas drag of the nebula. Moreover,  the capture radius of a protoplanet with a (tiny) envelope is also larger for small planetesimals. However, planets migrate as a result of disc-planet angular momentum exchange, with important consequences for
   their survival: due to the slow growth of a protoplanet in the oligarchic regime, rapid inward type I migration has important implications on intermediate mass planets that have not started yet their runaway accretion phase of gas. Most of these planets are lost in the central star. Surviving planets have either masses below 10 M$_\oplus$ or above several Jupiter masses. }
   { To form giant planets before the dissipation of the disc, small planetesimals ($\sim$ 0.1 km) have to be the major contributors of the solid accretion process.  However, the combination of  oligarchic growth and fast inward migration leads to the absence of intermediate mass planets. Other processes must therefore be at work in order to explain the population of extrasolar planets presently known. }


   \maketitle
%

\section{Introduction}

Since the discovery of the first extrasolar planet around a solar-type star (Mayor \& Queloz, 1995) about 800 extrasolar planets have been identified. Observations indicate that planets are abundant in the universe. Planets orbiting stars show a great variety of semi-major axis (from less than 0.01 AU to more than hundreds of AU) and masses (from less than an Earth mass to several Jupiter masses), as can be found in The Extrasolar Planets Encyclopaedia (http://exoplanet.eu/).   Planet formation models should be able to explain this observed diversity. The sequential accretion model or also called core nucleated accretion model is currently the most accepted scenario for planetary formation (e.g. Mizuno 1980, Pollack et al. 1996, Alibert et al. 2005 a, b, among others), as it can account naturally for the formation of planets in all mass ranges\footnote{Note, however, that the formation of planets at large distance from their central star seems very difficult in the core-accretion model.}.  It proposes that planetary growth occurs mainly in two stages. In the first stage,  the formation of  planets is dominated by the accretion of solids. If the protoplanet is able to grow massive enough ($\sim 10$ M$_\oplus$)  while the gas component of the protoplanetary disc is still present, it can bind gravitationally some of the surrounding gas, giving birth to a gas giant planet. The accretion of gas is slow at the beginning:  the planet growth is dominated by the accretion of solids and the energy released from the accreted planetesimals slows down the accretion of the envelope. When the accretion of planetesimals declines (generally because the protoplanet has emptied its feeding zone) the accretion of gas is triggered, and the planet can accrete hundreds of Earth masses in a very short time. The runaway accretion of gas characterises the second stage of the sequential accretion model, where the growth of the planet is dominated by the accretion of gas.

Since it was first proposed (Mizuno 1980), the sequential accretion model has been extensively studied and improved, trying to include the many fundamental processes that occur simultaneously with the growth of the planet, and that impact directly on it. Constructing a complete model that accounts for all these processes in a reasonable way is a hard task. Among its main ingredients, it has to include a realistic model for an evolving protoplanetary disc, a model for the accretion of solids and gas to form the planets (which itself requires a knowledge of the internal structure of the planet), a model for the interactions between the planets and the disc, and a model for the interactions between the forming planets. Each of these topics represent itself  an independent, ongoing area of research. Alibert et al. (2005) (A05 from now on) include some of these processes in a single planet formation model. Given the complexity of the problem or the unknowns related to some of these processes, many simplifications have to be assumed in order to keep the problem tractable from the physical and computational point of view. This is specially important, as we aim to compute thousands of simulations in order to account for the wide range of possible initial conditions (see Mordasini et al. 2009, M09 from now on, for more details). Therefore, our models represent a compromise between accuracy and simplicity in their physical description. 

The first stage of planetary formation corresponds to the growth of the solid embryo, which is dominated by the accretion of planetesimals. The growth 
of an embryo proceeds itself in two different regimes (Ida \& Makino 1993, Ormel et al. 2010). At the beginning, big planetesimals, that have larger cross-sections, 
are favoured to grow even bigger by accreting planetesimals that they encounter on their way. Being more massive, in turn, enlarges the gravitational focusing, which leads to accretion in a runaway fashion. 
However,  at some point, these runaway embryos become massive enough to stir the planetesimals around them. This results in an increase of  the relative velocities and the corresponding reduction of  the gravitational focusing. Growth among small planetesimals is stalled and only  big embryos have the possibility to continue accreting, although at a slower pace. This second regime in the growth of 
solid embryos is known as oligarchic growth, as only the larger planetesimals or embryos (the oligarchs) are able to keep on growing.  One important aspect is that the transition between runaway and  
oligarchic growth occurs for very small embryos.  As shown in Ormel et al. (2010), the actual mass for this transition depends upon many factors: size of the accreted planetesimals, location in the 
disc, surface density of solids, among others. In most of the cases,  an embryo of  $\sim$ 0.01 M$_\oplus$, or even smaller, is already growing in the oligarchic regime. 

Our model is building up on the model of A05 and M09. Our primary aim is to study the formation of  planets of different sizes,  and in particular the cases for which
 the accretion of gas is important. Therefore, in the computations presented here, we focus on the first phase of planetary formation, when the gas component of the disc
 is still present (in the case of small rocky planets, collisions between embryos after the 
 dissipation of the disc should be included in order to calculate their final masses).  For the formation of giant planets, the growth of the solid core is dominated by the oligarchic growth. 
 One of the weak points of the majority of previous giant planet formation models (e.g. Pollack et al. 1996, Hubickyj et al. 2005, A05, M09, Lissauer 2009,
 Mordasini et al. 2012, hereafter M12) is the description of the solid disc, in particular of the interactions between forming planets and planetesimals. Such simplified models lead to an overestimation of
 the solid accretion rate which, in turn, results in an underestimation of the formation time of the whole planet.

Indeed, in those works,  the model for the accretion rate of solids is oversimplified:  the whole 
formation of giant planets' cores is assumed to proceed very fast, underestimating the excitation that planetesimals suffer due to the presence of the embryos. When oligarchic growth is adopted as the dominant growth model, giant planet formation turns out to be more difficult. 
Formation times  become much longer than the typical lifetime of the protoplanetary disc. Fortier et al. (2007, 2009) and Benvenuto et al. (2009) studied the formation of giant planets adopting 
the oligarchic growth for the core. Assuming in situ formation for the planets and a simple, non evolving protoplanetary disc,  they show that the formation of giant planets is unlikely if planetesimals 
populating the disc are big (more than a few kilometres in size). However, formation could be accelerated if most of the accreted mass is in small planetesimals (less than 0.1 km). 
Guilera et al. (2010, 2011), also considering in situ models, studied the simultaneous formation of several planets where planetesimal drifting is included. They consider different density 
profiles for the disc and they find that, only in the cases of massive discs, the formation of the giant planets of the Solar System is possible only if planetesimal radii are smaller than 1 km. These models, 
however, do not take into account that planets would likely migrate during their formation.

In this work we include in our planet formation model  a more realistic description for the accretion of solids.  In Sect. \ref{sec:formationmodel},  we review the basics of the A05 formation 
model, presenting some improvements in the computation of the disc structure, internal structure and migration.  In Sect. \ref{accr_of_solids} we describe the new treatment 
of the  accretion of planetesimals.  In Sect. \ref{results} we present the results obtained for the formation of  isolated planets (the formation of planetary systems is described in Alibert et al. 2012,
and Carron et al. 2012). In Sect. \ref{discussion} we discuss our results and put them in context. Finally,  in Sect. \ref{conclusions} we summarise our results and underline the main conclusions.

\section{Formation model}
\label{sec:formationmodel}

The model and the numerical code used to calculate the formation of planets is in essence the same as in A05. In what follows, we  make a summary of the most relevant aspects 
of the model and the improvements that have been introduced since that work.  In the next section, we focus on the accretion rate of solids and we describe in detail the adopted 
model for the protoplanet-planetesimals interactions.

\subsection{Protoplanetary disc: gas phase}
\label{disc}

The structure and evolution of the protoplanetary disc is computed by first determining the vertical structure of the disc, for each
distance to the central star, and second, computing the radial evolution due to viscosity, photoevaporation, and mass accretion by forming planets.

\subsubsection{Vertical structure}

The vertical disc structure is
computed by solving the following equations:
\beq
{1 \over \rho_\mathrm{gas}}  \dpartial{P}{z} = - \Omega^2 z
\label{eq_disc_hydro}
,
\eeq
\beq
\dpartial{F}{z} = {9 \over 4} \,\rho_\mathrm{gas} \nu \Omega^2
\label{eq_disc_ener}
,
\eeq 
and 
\beq
F = {- 16 \pi \sigma_\mathrm{SB} T^3 \over 3 \kappa \rho_\mathrm{gas}} \dpartial{T}{z}
\label{eq_disc_diff}
.
\eeq
They reflect the hydrostatic equilibrium, the energy conservation, and the diffusion for the radiative flux. In these equations, 
$z$ is the vertical coordinate, $\rho_\mathrm{gas}$ the gas density, $P$ 
the pressure, $T$ the temperature,  $\nu$ the macroscopic viscosity,  $F$ the radiative flux, $\kappa$ is the opacity (Bell \& Lin 1994), and $\sigma_\mathrm{SB}$ is 
the Stefan-Boltzmann constant.  The Keplerian frequency, $\Omega$, is given by
\beq
\Omega^2 = G  M_{\star} / a^3,
\label{eq:omega}
\eeq
$G$ being the gravitational constant,
$M_{\star}$ the mass of the central star and $a$ the distance to the star\footnote{Note that we assume that the disc is thin, and the distance to the central
star does not vary on a vertical slide of the disc.}.

The equations (\ref{eq_disc_hydro})-(\ref{eq_disc_diff}) are solved with four boundary conditions. The first three are 
the temperature, the pressure and the energy flux at the surface. The surface of the disc is defined as the place where the vertical optical depth (between
the surface and infinity) is equal to 0.01. The fourth boundary condition is that the energy flux equals 0 in the midplane (see A05 for details).
The three differential equations, together with the four boundary conditions, have a solution only for one value of the disc thickness
$H$ which gives the location of the disc surface.
The macroscopic viscosity $\nu$ is calculated using the standard 
Shakura \& Sunyaev (1973) $\alpha-$parametrization, $\nu = \alpha c_s^2 / 
\Omega$.  The speed of sound $\cs$ is determined from the equation 
of state (Saumon et al. 1995). The temperature at the surface $\tsurf$ is computed as in A05. In the models
presented in this paper, $\alpha$ is set to 7$\times 10^{-3}$. This value of the alpha parameter has to be taken as an example.

In the calculations of this work we neglect irradiation and the possible presence of a dead zone. These effects will be included in future works. 

\subsubsection{Evolution}
\label{evolution}

The evolution of the gas disc surface density ($\Sigma = \int_{-H}^{H} \rho_\mathrm{gas} dz$) is computed by solving the diffusion equation:
\beq
{d \Sigma \over d t} = {3 \over a} {\partial \over \partial a } \left[ a^{1/2} {\partial \over \partial a}
\tilde{\nu} \Sigma a^{1/2} \right] + \dot{\Sigma}_w(a) + \dot{Q}_{\rm planet}(a),
\label{eqdiff}
\eeq
where $\tilde{\nu}$ is the effective viscosity, $\tilde{\nu} \equiv  \frac{1}{ \Sigma }\int_{-H}^{H} \nu \rho_\mathrm{gas} dz$ .
Photoevaporation is included using the model of Veras \& Armitage (2004):
\begin{eqnarray}
\left\{
\begin{array}{l}
\dot{\Sigma}_w  = 0  \,\,\,\,\,\,\,\,\,\,\, {\rm for}  \,\,\, a < R_g , \\
\dot{\Sigma}_w  \propto a^{-1}  \,\,\,\,\,\ {\rm for}  \,\,\, a > R_g ,
\end{array}
\right.
\end{eqnarray}
where $R_g=5$  AU. The total mass loss due to  photoevaporation is a free parameter. The sink term $\dot{Q}_{\rm planet}$ 
is equal to the gas mass accreted by the forming planets. For every forming planet, mass is removed from the protoplanetary disc in an 
annulus centred on the planet,  with a width equal to the planet's Hill radius 
\begin{eqnarray}
R_\mathrm{H}= a_M\left( \frac{M}{3{M_{\star}}}\right)^{1/3},
\label{eq:hillradius}
\end{eqnarray}
where $M$ is the total mass of the planet and $a_M$ is the location of the planet.

Eq. \ref{eqdiff} is solved on a grid which extends from the innermost radius of the disc to 1000 AU. At these two points,
the surface density is constantly equal to 0. The innermost radius of the disc is of the order of 0.1 AU.

\begin{figure}[h]
\center
\resizebox{\hsize}{!}{
\includegraphics{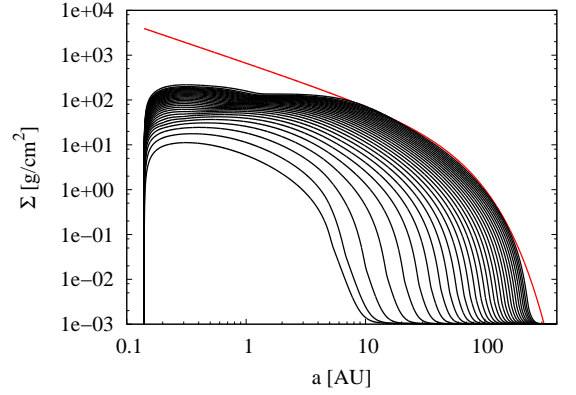}
 }
\caption{Disc model. Gas surface density as a function of radius at different times. For the
initial model (red line) the parameters are those of the first row of Table \ref{table_Andrews}, the disc lifetime being 3 Myr. 
Each line corresponds to a different time (from top to bottom 0 to 3 Myr, every 0.1 Myr). }
\label{evol_disc}
\end{figure}

Fig. \ref{evol_disc} presents a typical evolution of a disc, whose parameters correspond to the first row of table \ref{table_Andrews},
where the curves are plotted every $10^5$ years. In this model, the photoevaporation term is adjusted in order to obtain a 
disc lifetime equal to 3 Myr.

The characteristics of the protoplanetary disc  are chosen to match as close as possible the observations. The initial disc density profiles we consider are given by:
\beq
\Sigma = (2 - \gamma) { \mdisc \over 2 \pi \ac ^{2-\gamma} a_0^\gamma } \left( {a \over a_0} \right)^{-\gamma} \exp \left[ - \left( {a \over \ac} \right) ^{2-\gamma} \right]
,
\label{eq:init_sigma}
\eeq
where $a_0$ is equal to 5.2 AU.  The mass of the disc ($\mdisc$),  the characteristic scaling radius ($\ac$) and the power index  ($\gamma$) are derived from the observations of
Andrews et al. (2010). Adopting this kind of initial density profile is a difference from previous works (A05 and M09). For numerical reasons, the innermost disc radius, $a_\mathrm{inner}$, 
is always greater than or equal to 0.1 AU, and differs in some cases from the one cited in Andrews et al. (2010). The afore-mentioned parameters used to generate the initial disc's profile are listed in 
table \ref{table_Andrews}. Note that Andrews et al. (2010) derive also a value for the viscosity parameter $\alpha$. On the contrary, and for simplicity, we assume here that the viscosity parameter is the same for all
the protoplanetary discs we consider ($\alpha=7\times 10^{-3}$). Using a different $\alpha$ parameter will be the subject of future work. Note also that,  in the observations of 
Andrews et al. (2010), the mass of the central star ranges from 0.3 to 1.3 M$_\odot$. However, we assume here that these disc profiles are all suitable for protoplanetary discs around solar mass stars. 
Future disc observations will help improving this part of our models.

\begin{table}
\caption{Characteristics of disc models}
\begin{tabular}{lcccc}
\hline\noalign{\smallskip}
disc & $\mdisc$ ($M_\odot$) & $\ac$ (AU)  & $a_\mathrm{inner}$ (AU) & $\gamma$ \\
\noalign{\smallskip}
\hline\noalign{\smallskip}
1 & 0.029 & 46 & 0.14  & 0.9 \\
2 & 0.117 & 127 & 0.16  & 0.9 \\
3 & 0.143 &198 & 0.10 & 0.7 \\
4 & 0.028 & 126 & 0.10  & 0.4 \\
5 & 0.136 & 80 & 0.10  & 0.9 \\
6 & 0.077 & 153 & 0.12  & 1.0 \\
7 & 0.029 & 33 & 0.10  & 0.8 \\
8 & 0.004 & 20 & 0.10  & 0.8 \\
9 & 0.012 & 26 & 0.10  & 1.0 \\
10 & 0.007 & 26 & 0.10  & 1.1 \\
11 & 0.007 & 38 & 0.10  & 1.1 \\
12 & 0.011 & 14 & 0.10  & 0.8 \\
\hline
\end{tabular}
\label{table_Andrews}
\end{table}

As in A05, the planetesimal-to-gas ratio is assumed to scale with the metallicity of the central star. For every protoplanetary disc we consider, we  select at random the metallicity of a star from a list of 
$\sim 1000$ CORALIE targets (Santos, private communication).

Finally, following Mamajek (2009), we assume that the cumulative distribution of disc lifetimes decays exponentially with a characteristic time of 2.5 Myr. When a lifetime $\tdisc$ is 
selected, we adjust the photoevaporation rate in order that the protoplanetary disc mass reaches $10^{-5} \mathrm{M_\odot}$ at the time $t = \tdisc$, when we stop the calculation.

\subsection{Protoplanetary disc: solid phase}
\label{solid_disc}

We consider that the planetesimal disc is composed of rocky and icy planetesimals. Here we assume a mean density of 3.2 g/$\mathrm{cm^3}$ for the rocky ones,  and 1 g/$\mathrm{cm^3}$ for the icy. The rocky planetesimals are located between the innermost point of the disc (given by the fourth column of table \ref{table_Andrews}), and the initial location of the ice line, whereas the disc of icy planetesimals extends from the ice line to the outermost point in the simulation disc.

The location of the ice line is computed from the initial gas disc model, using the central temperature and pressure. The ice sublimation temperature we use depends upon the pressure. 
Note that in our model, the location of the ice line does not evolve with time. In particular, no condensation of moist gas, or sublimation of icy planetesimals is taken into account. Moreover,
the location of the ice line is based on the central pressure and temperature, meaning that the ice line is taken to be independent of the height in the disc. Note that in reality the ice line is likely to 
be an "ice surface" whose location depends upon the height inside the disc (see Min et al. 2011).

For the models presented here, we assume that all planetesimals have the same radius. Planetesimals' mass  is calculated assuming that they are spherical and have constant density (which depends 
on their location in the disc), and it   does not evolve with time. The extension of our calculations towards non-uniform and time evolving planetesimal mass function is something we are working on and will be included in the next paper.

The surface density of planetesimals, $\Sigma_m$, is assumed to be proportional to the initial surface density of gas $\Sigma_0$.  This means that
\begin{equation}
\Sigma_m (a) = f_\mathrm{D/G} f_\mathrm{R/I} (a) \Sigma_0 (a),
\end{equation}
where $f_\mathrm{D/G}$  is the dust-to-gas ratio of the disc and it scales with the metallicity of the central star (the lowest value being 0.003 and the largest one 0.125), and $f_\mathrm{R/I}$ takes into account the degree of  condensation of heavy elements. As in M09, we consider $f_\mathrm{R/I}= 1/4$ inside the ice line, and  $f_\mathrm{R/I}= 1$ beyond it, for rocky and icy planetesimals respectively.

The surface density of planetesimals evolves as a result of accretion and ejection by the forming planets. The same procedure as in A05 is adopted. Planetesimals that can be accreted 
by a growing planet are those within the planet's  feeding zone, here assumed to be an annulus of $5\,R_\mathrm{H}$ at each side of the orbit. The solids surface density inside the feeding zone 
is considered to be constant, i.e. the protoplanet  instantaneously homogenises it as a result of the scattering it produces among planetesimals. Ejected planetesimals are considered to be lost. 

We stress that this work is the first one of a series of papers. Here,  effects as planetesimal drifting due to gas drag, fragmentation and planetesimal size distribution are neglected, but will be included in future works.   

\subsection{Gas accretion: attached phase}
\label{equations}

\textit{Equations}

Planetary growth proceeds through solids and gas accretion. Gas accretion is a result of a planet's contraction, and is computed by solving the
standard internal structure equations:
\begin{eqnarray}
{d r^3 \over d M_r} & = & {3 \over 4 \pi \rho} \label{eq_r} , \\
{d P \over d M_r}&  = & {- G ( M_r + \mcore ) \over 4 \pi r^4} \label{eq_momentum}, \\
{d T \over d P}&  = & \nabla_{\rm ad} \,\,\, {\rm or} \,\,\, \nabla_{\rm rad} \label{eq_T} ,
\end{eqnarray}
where $r,P,T$ are respectively the radius, the pressure and the temperature inside the envelope. These three quantities depend upon the 
gas mass, $M_r$, included in a sphere of radius $r$. 
The temperature gradient is given by the  adiabatic ($\nabla_{\rm ad}$) or by the radiative gradient ($\nabla_{\rm rad}$),
depending upon the stability of the zone against convection, which we check using  the Schwarzschild criterion. These equations are solved using the equation of state (EOS)
of Saumon et al. (1995). The opacity, which enters in the radiative gradient $\nabla_{\rm rad}$ is computed from Bell and Lin (1994). In this work we assume that the grain opacity is the full 
interstellar opacity. However, Podolak (2003) and Movshovitz \& Podolak (2008) showed that the  grain opacity in the envelope of a forming planets should be much lower than the interstellar one. 
Reducing the grain opacity accelerates the formation of giant planets (Pollack at al. 1996, Hubickyj et al. 2005) because it allows the runaway of gas to start at smaller core masses. 
Since the objective of the present paper is to explore the consequences of the planet-planetesimal interactions, we only consider a single opacity, corresponding to the full interstellar opacity.

Note that we have omitted the energy equation that gives the luminosity, which itself enters in the radiative gradient (in the parts of the planet that are stable against convection), 
and in the determination of the planet stability to convective motions.  Including the energy equation in its standard form
 \mbox{${d L \over d M_r}  = \epsilon_\mathrm{m} - T {d S \over d t}$}  (the first term resulting from the accretion of planetesimals, the second one to the contraction of the planet) brings usually numerical difficulties.
 Here we follow Mordasini et al. (2012b)  to calculate the luminosity in an easier way. Note, however, that we improved their approach to take into account analytically the energy of the core.

The total luminosity is given by \mbox{$L= L_\mathrm{cont}+L_\mathrm{m}$,  $L_\mathrm{cont}$} being the 
contraction luminosity and $L_\mathrm{m}$ the accretion luminosity. We assume $L_\mathrm{cont}$  to be constant in the whole planetary envelope.  $L_\mathrm{cont}$ is computed  as the result of the change in total energy of the planet between two time steps $t$ and $t + dt$:
\beq
L_\mathrm{cont} = - { E_{\rm tot}(t+dt) - E_{\rm tot}(t) - E_{\rm gas, acc} \over d t } 
,
\eeq
where $E_{\rm tot}$ is the total planetary energy and \mbox{$E_{\rm gas,acc} = dt \dot{M}_{\rm gas} u_{\rm int}$}  is the energy advected during gas accretion ($u_{\rm int}$ being the internal 
specific energy). $E_{\rm gas,acc}$ is negligible compared to the other terms. The luminosity due to accretion of planetesimals is  
\beq
L_\mathrm{m}  = G { \dot{M}_{\rm core} \mcore \over \rcore }.
\eeq 

However, the energy at the time $t+dt$ is not known before computing the internal structure at this given time. To circumvent this problem, we use the following 
approach: the energy is split in two parts, one related to the core, one to the envelope. 

The core energy is given by \mbox{$E_{\rm core} = -(3/5) \, G M^2_{\rm core} / R_{\rm core}$}, the core density 
being assumed to be uniform. 
The envelope energy is assumed to follow a similar functional form: \mbox{$E_{\rm env} = - k_{\rm env} M_{\rm env} g$}, where $g$ is a mean gravity, taken to be
\mbox{$ G (M_{\rm core} / R_{\rm core} + \mtot / \rplanet)$}. This last formula defines $k_{\rm env}$, in which all our ignorance
of the internal structure is hidden. In order to calculate the envelope energy at time $t+dt$, the value of $k_{\rm env}$ is first taken to be the value resulting
from the structure at time $t$. Then, iteration on $k_{\rm env}$ is performed until convergence is reached. In general, only a first order correction is enough
to reach a satisfactory solution.

\textit{Boundary conditions}

The internal structure equations are solved for $M_r$ varying between the core mass $\mcore$, and the total planetary mass.
Four boundary conditions are given, namely the core radius,  the total planetary radius, and $\tsurfpla$ and $\psurfpla$, the temperature and 
pressure at this point. Given the boundary conditions, the differential equations have only one solution for a given 
total planetary mass.

The core radius is given as a function of the core luminosity and the pressure at the core surface by the following
formula, which constitute a fit to the results of Valencia et al. (2010):
\begin{eqnarray}
{ \rcore \over 9800 {\rm km} } &=&  \left( {\mcore \over 5 \mearth} \right) ^{0.28+0.02 \sqrt{ {\mcore \over 5 \mearth}}} \times \nonumber\\
&&10^{- \left[ \log_{10} \left(1 +  {\pcore / \sqrt{ \mcore \over 5 \mearth} }) \right) / 7 \right] ^3} 
,
\end{eqnarray}
where $\pcore$ is the pressure a the core-envelope interface, expressed in GPa.

The radius of the planet, $R_M$, is given by (Lissauer et al. 2009):
\beq
R_M = { G M \over \frac{\cs^2}{k_1} + {G M \over k_2 \rhill }}
, 
\eeq
where $\cs^2$ is the square of the sound velocity in the disc midplane at the planet's location, $k_1= 1$ and $k_2= 1/4$.
At the planet's surface, the temperature and pressure are given by:
\beq
\tsurfpla = \left( \tdisc ^4 + { 3 \tauout L \over 16 \pi \sigma_P R_M^2 } \right)^{1/4}  
,
\eeq
and
\beq
\psurfpla = \pdisc
,
\eeq
with  $\tauout = \kappa(\tdisc,\rhodisc) \rhodisc R_M$, $L$ is the planet luminosity, and  $\tdisc$, $\rhodisc$, $\pdisc$ are the temperature, density and pressure in
the disc midplane at the location of the planet.

\subsubsection{Gas accretion: detached phase}

By solving the differential equations (\ref{eq_r}) to (\ref{eq_T}) with the boundary conditions mentioned above, one can derive the planetary envelope mass as a function of time, and therefore the gas
accretion rate $\dot{M}_{\rm gas}$. However, the rate of gas accretion that can be sustained by the protoplanetary disc is not arbitrary, and is in particular limited by the viscosity. When the gas
accretion rate required by the forming planet is larger than the one that can be delivered by the disc,  $\mdotgasmax$, the planet goes into the detached phase.

In the detached phase, the planetary growth rate by gas accretion does not depend upon its internal structure, but is rather given by the structure and evolution of the disc. During this phase, the
internal structure is given by solving the same equations (\ref{eq_r}) to (\ref{eq_T}), this time for a  mass $M_r$ ranging from $\mcore$ to $\mplanet$ (which is known). The boundary conditions are
the same, except two of them:
\begin{itemize}
\item the pressure, that  includes the dynamical pressure due to gas free falling from the disc to the planet,
\beq
\psurfpla = \pdisc + { \dot{M}_{\rm gas} \over 4 \pi R_M^2 } \vff
.
\eeq
In this equation,  $\vff$ is the free falling velocity from the Hill radius to the planetary radius,  \mbox{$\vff = - \sqrt{2 G M \times \left( 1/R_M - 1/\rhill \right)}$}.
Note that  the planetary radius is not known \textit{a priori}, but  computed as a result of integrating
Eqs. (\ref{eq_r}) to (\ref{eq_T}).

\item the maximum accretion rate, $\mdotgasmax$, which is equal to 
\beq
\mdotgasmax = \max \left[ F \left( a_M+\rhill \right), 0 \right] +  \min \left[ F \left(a_M-\rhill \right), 0 \right]  
\label{maxdotgas} 
\eeq
where \mbox{$F = 3 \pi \nu \Sigma + 6 \pi r \dpartial{\nu \Sigma}{a} $} is the mass flux in the disc. Geometrically, the maximum accretion rate that can be provided by the disc is equal 
to the mass flux entering the planet's gas feeding zone.  The gas can enter  either from the outer parts of the disc (which is the general case), or from the inner part of the disc (which 
can be the case in the outer part of the disc).
\end{itemize}

Therefore, during a time step, a planet has access to the mass delivered at its location by the disc ($\mdotgasmax \times dt$), and
a mass reservoir made of the gas mass already present in the planet's gas feeding zone (see also Ida and Lin 2004). This reservoir of 
gas is assumed to be empty when the planet is massive enough to open a gap (which coincides with the transition to type II migration, see next section).  However, the feeding zone continues receiving gas due to viscosity at the local accretion rate (Eq. \ref{maxdotgas}).

\subsection{Orbital evolution: disc-planet interaction}
\label{orbital_evolution}

Disc-planet interaction leads to planet migration, which can occur in different regimes. For low mass planets, not massive enough to open a gap in the protoplanetary disc, 
migration occurs in type I (Ward 1997, Tanaka et al.  2002, Paardekooper et al.  2010,  2011). For higher mass planets, migration is again sub-divided in two modes: disc-dominated type II 
migration, when the local disc mass is larger than the planetary mass (the migration rate is then simply given by the viscous evolution of the protoplanetary disc), and planet-dominated type II 
migration in the opposite case (see M09). The transition between type I and type II migration occurs when 
\beq
{3 \over 4} { \hdisc \over \rhill} + {50 M_\star \over M \mathrm{Re} } = 1,
\eeq
(Crida et al.  2006), where $\hdisc$ is the disc scale-height at the location of the planet, and 
$\mathrm{Re} = { a_M^2 \Omega \over \nu }$ is the macroscopic Reynolds number at the location of the planet ($\nu$ is the same as the one used for the disc evolution). 

First models of type I migration (Ward 1997, Tanaka et al. 2002) predicted so rapid migration rates that it was necessary to reduce arbitrarily the migration rate by a constant factor, named $\f1$ in A05 and M09, in order to reproduce
observations. Since these first calculations, type I migration has been studied in great details, and new formulations for type I migration rates are now available (Paardekooper et al.  2010, 2011). We use in our model 
an analytic description of type I migration, which reproduces the results of Paardekooper et al. (2011). A detailed description of this model is presented in Dittkrist et al. (in prep.), and preliminary results have been 
presented in Mordasini et al. (2010).

\section{The accretion rate of solids}
\label{accr_of_solids}

The growth of the solid component of a protoplanet, $M_\mathrm{core}$,  is assumed to be due to the accretion of planetesimals. Adopting the particle-in-a-box approximation, its growth rate can be calculated with 
\begin{eqnarray}
\frac{\mathrm{d}M_\mathrm{core}}{\mathrm{d}t}=  \left( \frac{2\pi \Sigma_m R_\mathrm{H}^2}{P_{\mathrm{orbital}}}\right) ~P_\mathrm{coll},
\label{eq:mdotcore}
\end{eqnarray}
(Chambers 2006), where $\Sigma_m$ is the  solids surface density  at the location of the protoplanet and $P_{\mathrm{orbital}}$ is its orbital period. The collision rate, $P_\mathrm{coll}$, is the probability 
 that a planetesimal is accreted by the protoplanet. This probability depends upon the relative velocity between planetesimals and the protoplanet which, in turn, depends upon the planetesimals 
eccentricities and inclinations. We represent by $e$ ($i$) the root mean square of the eccentricity (inclination) of planetesimals.  Planetesimals are found to be in different velocity regimes depending on their 
random velocities. These regimes are known as high-, medium- and low- velocity  regime. Each regime is characterised by a range of values of planetesimal's  reduced eccentricities ($\tilde{e}=ae/R_\mathrm{H}$) 
and inclinations  ($\tilde{i}=ai/R_\mathrm{H}$):  the high-velocity regime is defined by $\tilde{e}, \tilde{i} \gtrsim 2$, the medium-velocity regime by  $2\gtrsim \tilde{e}, \tilde{i} \gtrsim 0.2$  and the low-velocity regime by $\tilde{e}, \tilde{i} \lesssim 2$. This leads to different collision rates 
\begin{eqnarray}
P_\mathrm{high}&=& \frac{(R+r_m)^2}{2\pi R_\mathrm{H}^2} \left(I_\mathrm{F}(\beta)+\frac{6R_\mathrm{H}I_\mathrm{G}(\beta)}{(R+r_m)\tilde{e}^2}\right), \label{eq:collisionprobability1}\\
P_\mathrm{med}&=& \frac{(R+r_m)^2}{4\pi R_\mathrm{H}^2 \tilde{i}} \left(17.3+\frac{232R_\mathrm{H}}{R+r_m} \right), \label{eq:collisionprobability2}\\
P_\mathrm{low}&=&11.3 \left(\frac{R+r_m}{R_\mathrm{H}}\right)^{1/2}, \label{eq:collisionprobability3}
\end{eqnarray}
(see Inaba et al. 2001 and references therein) where $R$ is the radius of the protoplanet (in the case of a solid body without a gaseous envelope $R$ is its geometrical radius), $r_m$ is 
the radius of the planetesimals, $\beta=\tilde{i}/ \tilde{e}$, and the functions $I_F$ and $I_G$ are well approximated by
\begin{eqnarray}
I_\mathrm{F}(\beta) &\simeq& \frac{1+0.95925\beta+0.77251\beta^2}{\beta(0.13142+0.12295\beta)},\label{eq:ifigfunctions1}\\
I_\mathrm{G}(\beta) &\simeq& \frac{1+0.3996\beta}{\beta(0.0369+0.048333\beta + 0.006874\beta^2)}, 
\label{eq:ifigfunctions}
\end{eqnarray}
for $0<\beta\leq 1$, which is the range of interest for this work  (Chambers 2006).

According to Inaba et al. (2001) the mean collision rate can be approximated by 
\begin{equation}
P_\mathrm{coll}= \mathrm{min}\left( P_\mathrm{med}, (P_\mathrm{high}^{-2}+P_\mathrm{low}^{-2})^{-1/2}\right).
\label{eq:collisionprobability}
\end{equation}

When an embryo is able to gravitationally bind gas from its surroundings it becomes more difficult to define its radius, which is not just the core radius. For the purpose of the collision rate,  the capture radius of the protoplanet should  
depend  upon the mass of the protoplanet,  upon the planetesimals' velocity with respect to the protoplanet, upon the density profile of the envelope, $\rho(r)$, and  upon the size of the accreted planetesimals
(smaller planetesimals are more affected by the gas drag of the envelope and therefore are easier to capture). As in Guilera et al. (2010), here we adopt the prescription of Inaba \& Ikoma (2003) where 
 the capture radius $R$ can be obtained by solving the following equation  

\begin{equation}
r_m=\frac{3}{2} \frac{\rho(R) R}{\rho_m} \left(\frac{v_\mathrm{rel}^2+2GM(R)/R}{\,v_\mathrm{rel}^2+2GM(R)/R_\mathrm{H}}\right),
\end{equation}
where $\rho_m$ is the planetesimals' bulk density, $G$ is the gravitational constant and the relative velocity $v_\mathrm{rel}$ is given by
\begin{equation}
 v_{\mathrm{rel}} = v_\mathrm{k} \sqrt{5/8 \, e^2+1/2\, i^2},
 \end{equation}
with $v_\mathrm{k}$ is the keplerian velocity ($v_\mathrm{k}=\Omega a$).  This simple formula for the capture radius approximates well more complex models (as the one described in A05) 
with the advantage that it reduces the computational time.

It is clear from the above equations that the accretion rate of solids depends upon the eccentricities and inclinations of planetesimals, which define their relative velocities with respect to the embryo: 
the higher the relative velocity is, the less likely planetesimals are captured by the embryo.  The eccentricities and inclinations of planetesimals are affected by the damping produced by the nebular 
gas drag, by the gravitational stirring of the protoplanet (protoplanet-planetesimal interactions) and, to a lesser extent, by their mutual gravitational interactions (planetesimal-planetesimal interactions):  
\begin{eqnarray}
\frac{\mathrm{d}e^2}{\mathrm{d}t}&=&\frac{\mathrm{d}e^2}{\mathrm{d}t}\bigg|_{\mathrm{drag}}+\frac{\mathrm{d}e^2}{\mathrm{d}t}\bigg|_{\mathrm{VS,}M}+\frac{\mathrm{d}e^2}{\mathrm{d}t}\bigg|_{\mathrm{VS,}m}, \label{eq:diff_eqe}\\
\frac{\mathrm{d}i^2}{\mathrm{d}t}&=&\frac{\mathrm{d}i^2}{\mathrm{d}t}\bigg|_{\mathrm{drag}}+\frac{\mathrm{d}i^2}{\mathrm{d}t}\bigg|_{\mathrm{VS,}M}+\frac{\mathrm{d}i^2}{\mathrm{d}t}\bigg|_{\mathrm{VS,}m}.
\label{eq:diff_eqi}
\end{eqnarray}
The first term represents the effect of the nebular gas drag, the second term the viscous stirring produced by an embryo of mass $M$ and the third term the planetesimal-planetesimal viscous stirring. 

The drag force experienced by a spherical body depends upon its relative velocity with respect to the  gas.  If we consider that the protoplanetary nebula is mainly composed by $\mathrm{H_2}$  molecules, 
the mean free path of a  molecule of gas is
\begin{equation} 
\lambda=(n_\mathrm{H_2} \sigma_\mathrm{H_2})^{-1}, 
\label{eq:lambda}
\end{equation}
where $n_\mathrm{H_2} $ is the number density of $\mathrm{H_2}$ molecules and $\sigma_\mathrm{H_2}$ is the collision cross-section of an $\mathrm{H_2}$ molecule.
Depending upon the ratio between the planetesimal's radius and the mean free path of the molecules,  three drag regimes can be defined (Rafikov 2004 and references therein). The first 
two drag regimes are for planetesimals which radii are larger than the mean free path, $r_m \gtrsim \lambda$. These are the quadratic and the Stokes regime. For the distinction of these regimes 
we adopt the criterion proposed by Rafikov (2004) in terms of the molecular Reynolds number $\mathrm{Re}_\mathrm{mol} \equiv v_{\mathrm{rel}} r_m/\nu_\mathrm{mol}$, where 
$\nu_\mathrm{mol}$ is the molecular viscosity, $\nu_\mathrm{mol} = \lambda c_\mathrm{s}/3$. If $\mathrm{Re}_\mathrm{mol}\gtrsim20$ we assume that the gas drag is in the quadratic regime, 
and the differential equations for the evolution of the eccentricity and inclination are given by 
\begin{eqnarray}
\frac{\mathrm{d}e^2}{\mathrm{d}t}\bigg|_{\mathrm{drag}}&=& -\frac{2e^2}{\tau_{\mathrm{drag}}} \left(\frac{9}{4}\eta^2+\frac{9}{4\pi}\xi^2 e^2+\frac{1}{\pi}i^2\right)^{1/2}, \label{eq:drag_quadratice}\\
\frac{\mathrm{d}i^2}{\mathrm{d}t}\bigg|_{\mathrm{drag}}&=& -\frac{i^2}{\tau_{\mathrm{drag}}} \left(\eta^2+\frac{\xi^2}{\pi}e^2+\frac{4}{\pi}i^2\right)^{1/2}, 
\label{eq:drag_quadratici}
\end{eqnarray}
(Adachi et al. 1976 corrected by Inaba et al. 2001)\footnote{We did not find any noticeable difference in our results when using the original formulas of Adachi et al. (1976) or the corrected ones by Inaba et al. (2001).}, with \mbox{$\xi \simeq 1.211$}. The value of $\eta$ depends upon the distance to the star, on the gas density and on the pressure gradient, $\mathrm{d}P/ \mathrm{d}a$,
\begin{equation}
\eta(a)=-\frac{1}{2\Omega^2 a \rho_\mathrm{gas}} \frac{\mathrm{d}P}{\mathrm{d}a}, 
\label{eq:eta}
\end{equation}
where $ \rho_\mathrm{gas}$ and $\mathrm{d}P/ \mathrm{d}a$ are derived from the disc model. 

The gas drag timescale is
\begin{equation}
\tau_\mathrm{drag}=\frac{8\rho_m r_m}{3C_\mathrm{D}\rho_\mathrm{gas}v_\mathrm{k}},
\label{eq:tau_drag}
\end{equation}
where $C_\mathrm{D}$ is the drag coefficient, which is of the order of unity.

The Stokes regime occurs for  $r_m \gtrsim \lambda$ and $\mathrm{Re}_\mathrm{mol}<20$, and the equations for the eccentricity and inclinations of planetesimals are 
\begin{eqnarray}
\frac{\mathrm{d}e^2}{\mathrm{d}t}\bigg|_{\mathrm{drag}}&=& -\frac{3}{2} \frac{\lambda c_\mathrm{s} \rho_\mathrm{gas} e^2}{\rho_m r_m^2}, \label{eq:drag_stokese}\\
\frac{\mathrm{d}i^2}{\mathrm{d}t}\bigg|_{\mathrm{drag}}&=& -\frac{3}{4} \frac{\lambda c_\mathrm{s} \rho_\mathrm{gas} i^2}{\rho_m r_m^2},
\label{eq:drag_stokesi}
\end{eqnarray}
(Adachi et al. 1976, Rafikov 2004).

Note that in this paper, as for example in Stepinski \& Valageas (1996),  we defined two different Reynolds numbers, $\mathrm{Re}$ and $\mathrm{Re}_\mathrm{mol}$, and two different viscosities, $\nu$ and $\nu_\mathrm{mol}$. 
The macroscopic quantities ($\mathrm{Re}$ and $\nu$) are a measure of the fluid dynamics of the disc in a global scale (to compute the evolution of the disc) while the microscopic
 quantities ($\mathrm{Re}_\mathrm{mol}$ and $\nu_\mathrm{mol}$) characterise the local state of the gas and are used to calculate the eccentricities and inclinations of planetesimals.  

When $r_m \lesssim \lambda$, the third regime, Epstein regime, takes place, and eccentricities' and inclinations' evolution follow the equations  
\begin{eqnarray}
\frac{\mathrm{d}e^2}{\mathrm{d}t}\bigg|_{\mathrm{drag}}&=& - e^2\frac{ c_\mathrm{s} \rho_\mathrm{gas} }{\rho_m r_m}, \label{eq:drag_epsteine}\\
\frac{\mathrm{d}i^2}{\mathrm{d}t}\bigg|_{\mathrm{drag}}&=& -\frac{i^2}{2} \frac{ c_\mathrm{s} \rho_\mathrm{gas}}{\rho_m r_m}.
\label{eq:drag_epsteini}
\end{eqnarray}
(Adachi et al. 1976, Rafikov 2004). In this paper we consider that the population of planetesimals is represented by spherical bodies of a single size. It is worth mentioning that, although we allow for the three drag regimes according to the above-mentioned criterions, for the ranges 
of planetesimal sizes considered in this work (100-0.1 km) and for the kind of interaction we are mostly interested in here (protoplanet-planetesimal interactions), planetesimals are found to 
be mainly in the quadratic regime. Therefore, in most of the cases, for determining the solids accretion rate the effect of the gas drag is governed by Eqs. (\ref{eq:drag_quadratice}) - (\ref{eq:drag_quadratici}).

Planetesimals' eccentricities and inclinations are excited by the presence of a protoplanet. Ohtsuki et al. (2002) studied the evolution of the mean square orbital eccentricities and inclinations and introduced semi-analytical formulae to describe the stirring produced by the protoplanet 

\begin{eqnarray}
\frac{\mathrm{d}e^2}{\mathrm{d}t}\bigg|_{\mathrm{VS,}M}&=& \left(\frac{M}{3bM_\star P_\mathrm{orbital}}\right) P_\mathrm{VS}, \label{eq:VSMe}\\
\frac{\mathrm{d}i^2}{\mathrm{d}t}\bigg|_{\mathrm{VS,}M}&=& \left(\frac{M}{3bM_\star P_\mathrm{orbital}}\right) Q_\mathrm{VS},
\label{eq:VSMi}
\end{eqnarray}
where $b$ is the full width of the feeding zone of the protoplanet in terms of their Hill radii (here we adopt $b\sim 10$), and $P_\mathrm{VS}$ and $Q_\mathrm{VS}$ are given by
\begin{eqnarray}
P_\mathrm{VS}&=& \left[ \frac{73 \tilde{e}^2}{10\Lambda^2}\right] \ln(1+10\Lambda^2/\tilde{e}^2)+\nonumber\\
&&+\left[\frac{72 I_\mathrm{PVS}(\beta)}{\pi \tilde{e} \tilde{i}}\right] \ln (1+ \Lambda^2),\label{eq:PVS} \\
Q_\mathrm{VS}&=& \left[ \frac{4\tilde{i}^2+0.2\tilde{i}\tilde{e}^3}{10\Lambda^2\tilde{e}}\right] \ln(1+10\Lambda^2\tilde{e}^2)+\nonumber\\
&&+\left[\frac{72 I_\mathrm{QVS}(\beta)}{\pi \tilde{e} \tilde{i}}\right] \ln (1+ \Lambda^2),
\label{eq:PVS}
\end{eqnarray}
with $\Lambda=\tilde{i}(\tilde{e}^2+\tilde{i}^2)/12$.  The functions $I_\mathrm{PVS}(\beta) $ and $I_\mathrm{QVS}(\beta)$ can be approximated, for $0<\beta\leq 1$, by
\begin{eqnarray}
I_\mathrm{PVS}(\beta) &\simeq& \frac{\beta-0.36251}{0.061547+0.16112\beta+0.054473\beta^2},\label{eq:QVSfunctions}\\
I_\mathrm{QVS}(\beta) &\simeq& \frac{0.71946-\beta}{0.21239+0.49764\beta + 0.14369\beta^2},
\label{eq:iPVSfunctions}
\end{eqnarray}
(Chambers 2006). The excitation that the protoplanet produces on the planetesimals weakens with the increase in the distance between the  protoplanet and the planetesimals, i.e.
further away planetesimals are less excited. Here we follow the approach of Guilera et al. (2010) and consider that the effective stirring is given by 

\begin{eqnarray}
\frac{\mathrm{d}e^2}{\mathrm{d}t}\bigg|_{\mathrm{VS,}M}^\mathrm{eff}&=& f(\Delta) \frac{\mathrm{d}e^2}{\mathrm{d}t}\bigg|_{\mathrm{VS,}M}, \label{eq:e_eff}\\
\frac{\mathrm{d}i^2}{\mathrm{d}t}\bigg|_{\mathrm{VS,}M}^\mathrm{eff}&=&  f(\Delta) \frac{\mathrm{d}i^2}{\mathrm{d}t}\bigg|_{\mathrm{VS,}M},
\label{eq:i_eff}
\end{eqnarray}
where $f(\Delta)$ ensures that the perturbation of the protoplanet is confined to its neighbourhood,
\begin{equation}
f(\Delta)=\left[1+\left(\frac{\Delta}{nR_\mathrm{H}}\right)^5\right]^{-1},
\end{equation} 
with $\Delta=|a_M-a_m|$, where $a_M$ is the semi-major axis of the protoplanet and  $a_m$ is the semi-major axis of the planetesimal. Although the functional form is arbitrary, the scale 
on which the stirring acts is similar to the one found in $N$-body calculations (excluding the effects of resonances). For this work we have chosen $n=5$ to limit the perturbation of the planet to
 its feeding zone. In the future, with the aid of $N$-body calculations,  we plan to have a better semi-analytical function to characterise the extent of the planetary perturbation.  

We also consider that planetesimals' eccentricities and inclinations are stirred by their mutual interactions. For a population of planetesimals of equal mass $m$, the evolution of 
their eccentricities and inclinations are well described by 
\begin{eqnarray}
\frac{\mathrm{d}e^2}{\mathrm{d}t}\bigg|_{\mathrm{VS,}m}&=& \frac{1}{6} \sqrt{\frac{Ga}{M_\star}}\Sigma_m h_m P_\mathrm{VS},\label{eq:psimal_eqe}\\
\frac{\mathrm{d}i^2}{\mathrm{d}t}\bigg|_{\mathrm{VS,}m}&=& \frac{1}{6} \sqrt{\frac{Ga}{M_\star}}\Sigma_m h_m Q_\mathrm{VS},
\label{eq:psimal_eqi}
\end{eqnarray}
(Ohtsuki et al. 2002), with 
\begin{equation}
h_m= \left(\frac{2m}{3M_\star}\right)^{1/3}.
\end{equation}
In this case $P_\mathrm{VS}$ and $Q_\mathrm{VS}$ are evaluated with the reduced eccentricity and inclination relative to the planetesimal mass (i.e. $\tilde{e}= 2e/h_m$, $\tilde{i}= 2i/h_m$). 
Note that there is no  dynamical friction term in Eqs. (\ref{eq:psimal_eqe}) - (\ref{eq:psimal_eqi}), as it vanishes when a single mass population of planetesimals is considered (Ohtsuki et al. 2002). 
 Although, strictly speaking we have two populations of planetesimals (rocky and icy bodies depending if they are inside or beyond the ice line) that have the same size but not the same mass 
 (because of the difference in their density), the region where the two types of planetesimals are present at the same time is very narrow. 
 
 In this work we neglect changes in the mass of planetesimals due to fragmentation and changes in the surface density owe to planetesimal drifting.
 
 \subsection{Comparison with the previous solids accretion rate} 
 
\begin{figure*}
\centering
 \includegraphics[width= 0.5\textwidth, angle=-90]{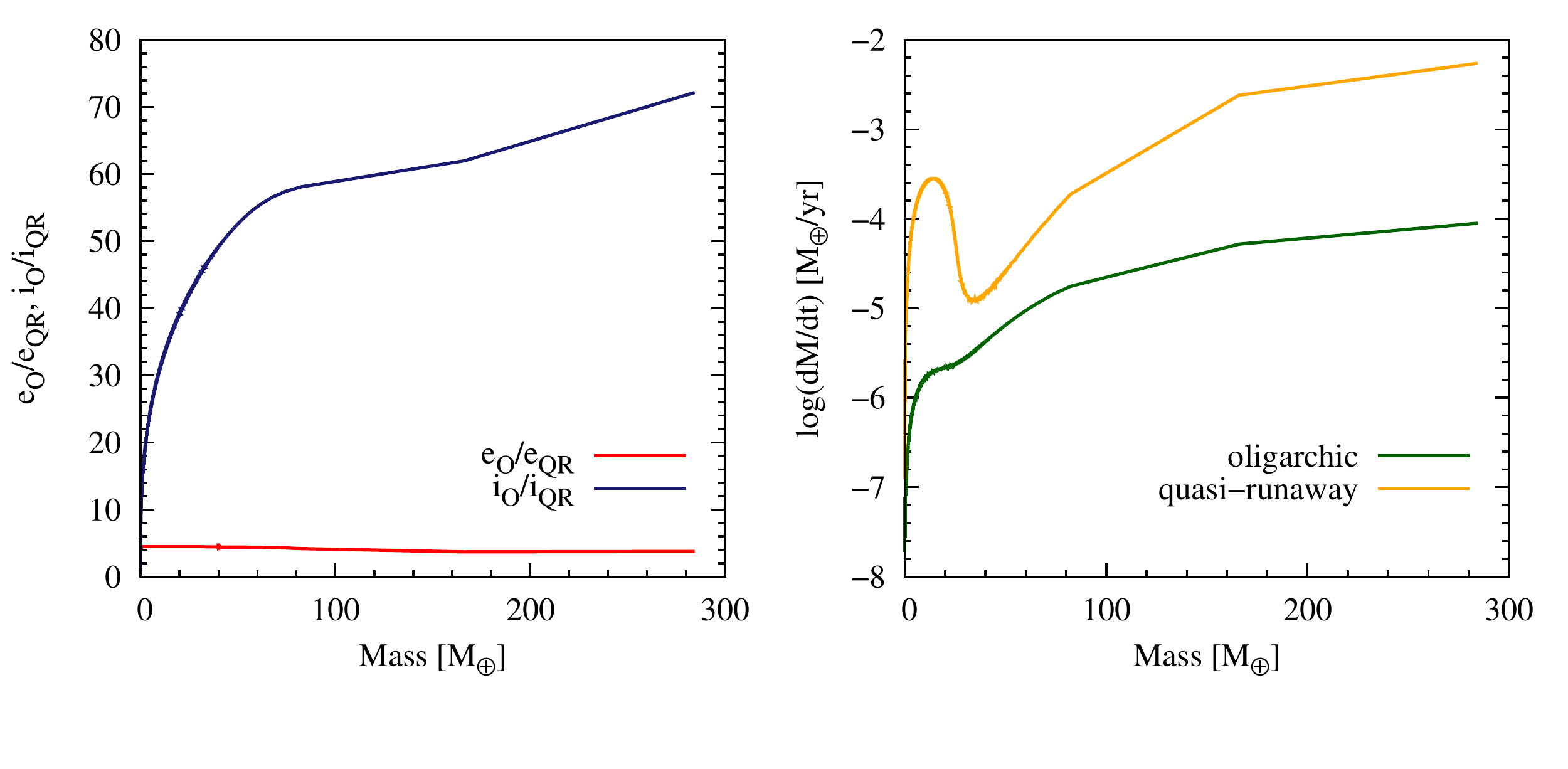} 

   \caption{Left panel shows the ratio between eccentricities (red) and inclinations (blue) as a function of the planet's mass considering both the oligarchic and the quasi runaway growth. Clearly, planetesimals excitation is several times higher in the oligarchic growth than in the quasi-runaway growth. Right panel shows the impact of planetesimals excitation in the accretion rate of solids: in the case of the quasi-runaway growth (yellow) the accretion rate is much larger than in the oligarchic growth (green).}
    \label{Fig:comp}
    \end{figure*}

 In previous works (A05, M09), the solids accretion rate has been treated in a very simple way leading to an underestimation of the formation timescale of planets. In those works, the prescription for planetesimals' eccentricities and inclinations was the same as in Pollack et al. (1996), were it was assumed that planetesimals' inclinations  depend only on planetesimal-planetesimal interactions. Under this assumption, the reduced value of planetesimals' inclination,  $\tilde {i}$, was prescribed as:
 
\begin{equation}
\label{eq:ihp96}
\tilde{i}= \frac{v_{\mathrm{E}}}{\sqrt{3} \Omega R_\mathrm{H}}, 
\end{equation}
where $v_{\mathrm{E}}$ is the escape velocity from the surface of a 
planetesimal. This means that  planetesimal's inclination $i= \tilde{i} \; R_\mathrm{H}/a$ 
is constant independently of the mass of the planet.  On the other hand,  eccentricities are assumed to be controlled by both planetesimals and protoplanet
stirring, its value given by: 
\begin{equation}
\label{eq:ehp96} 
\tilde{e}= \mathrm{max} (2 \tilde{i},2). 
\end{equation}
Therefore,  if 
$\tilde{e}=2 \tilde{i}$ the
protoplanet is growing according to the runaway regime because $e$ and $i$ would be 
independent of
the mass of the protoplanet.  If $\tilde{e}=2$, the eccentricity of
planetesimals would be affected by the presence of the protoplanet, so to a certain extent the  stirring of the embryo is taken into account. However, this
condition corresponds to the 
protoplanet--planetesimal scattering in the shear--dominated regime. Ida \& Makino (1993) showed that the shear--dominated regime lasts 
for only a few thousand years, after which  planetesimals are strongly 
stirred by the protoplanet. During the shear--dominated period,  eccentricities and
inclinations of planetesimals in the vicinity of the protoplanet remain low.
This leads to an accretion scenario which is much faster than that corresponding
 to the oligarchic regime (the oligarchic regime usually occurs in the dispersion--dominated regime). To show clearly the difference between this, let's call it  ``quasi-runaway'' accretion of solids, and the oligarchic regime, we performed two simulations that are identical in all parameters except for the prescription of $e$ and $i$. The planet is assumed to form in situ at 6 AU. Accreted planetesimals are 100 km in radius.  No disc evolution is considered, so simulations are stopped when the planet reaches one Jupiter mass. For the quasi-runaway regime we use Eqs. (\ref{eq:ehp96}) and (\ref{eq:ihp96}) to calculate $e$ and $i$, while for the oligarchic growth we solve the differential equations presented in the previous section. Fig. \ref{Fig:comp} (left panel) shows the ratio of oligarchic and quasi-runaway  eccentricities and inclinations as a function of the mass of the planet. In the case of the eccentricity, the one corresponding to  the oligarchic regime is $\sim$ 4 times larger than in the quasi runaway regime. The oligarchic inclination is several tens of times bigger than the quasi-runaway one. As a consequence, the accretion rate of solids is much smaller in the oligarchic than in the quasi runaway regime because in the oligarchic regime planetesimals are more exited and are more difficult to accrete. A comparison between the two accretion rates of solids is shown in Fig. \ref{Fig:comp} (right panel). Due to the smaller accretion rate in the oligarchic regime,  while in the quasi runaway regime it takes less than 1 Myr to form the planet, in the oligarchic regime formation is much longer, taking  $3.25\times 10^{7}$ years.

\section{Results}
\label{results}

In the previous section we introduced the main characteristics of the planet formation model, with special focus on the differences in the physical and numerical model  with respect to A05. 
In the following sections we will concentrate on the impact that the accretion rate of solids has on the formation of giant planets. Here we will consider the formation of isolated planets, i.e. only one planet per disc. 
 The computations of planetary system formation will be presented in other papers (Alibert et al. 2012, Carron et al. 2012).

As we described in Sect. \ref{accr_of_solids}, the treatment of the evolution of eccentricities and inclinations of planetesimals intends to minimise the assumptions on their values 
(keeping in mind that it is not an $N$-body calculation, but the adopted formulas reproduce $N$-body results of planetesimals accretion rates and excitation). 
In order to consider a realistic accretion rate but not too computationally expensive, Thommes et al. (2003) considered that planetesimals' eccentricities and inclinations can be estimated assuming 
that the stirring produced by the protoplanet is instantaneously balanced by the gas drag. The approximation to the  equilibrium values of $e$ ($e_\mathrm{eq}$) 
can be derived by equating the stirring timescale and the damping timescale, resulting in 
\begin{equation}
e_\mathrm{eq}=1.7\frac{ m^{1/15}M^{1/3}\rho_m^{2/15}}{b^{1/5}C_\mathrm{D}^{1/5}\rho_\mathrm{gas}^{1/5}M_{\star}^{1/3}a^{1/5}}.
\label{eq:e_eq}
\end{equation}

The equilibrium value for $i$ ($i_\mathrm{eq}$) is assumed to be half the value of $e_\mathrm{eq}$, as this relationship has been shown to be a good approximation in the high-velocity cases (Ohtsuki et al. 2002):
\begin{equation}
i_\mathrm{eq}=\frac{1}{2}e_\mathrm{eq}.
\label{eq:i_eq}
\end{equation}

However, it is not clear whether planetesimals are always in equilibrium, especially if we consider that depending on their mass, planetesimals are differently affected by gas drag, and that 
during its formation, a planet migrates and the protoplanetary disc evolves. On the other hand, if equilibrium is attained, it is interesting to compare the equilibrium values obtained by solving 
explicitly Eqs. (\ref{eq:diff_eqe})-(\ref{eq:diff_eqi}) with the approximations given by Eqs. (\ref{eq:e_eq})-(\ref{eq:i_eq}) based on timescales considerations.

Fortier at al. (2007, 2009) and Benvenuto et al. (2009) assume, for simplicity,  the equilibrium approximation of Thommes et al. (2003) in their in situ, giant planet formation models. However, 
Chambers (2006) finds that important deviations from equilibrium occur at the very beginning of the growth of the embryo but eventually equilibrium is attained.  In the cases he shows, these 
deviations do not seem to have a noticeable effect in the final mass of the planet as long as no time restriction for the lifetime of the disc is assumed.  On the other hand, Guilera et al. (2010, 2011) 
do not use any approximations and  calculate  explicitly   $e$ and $i$ by solving the corresponding time evolution differential equations  in their giant planet formation model.  
No study comparing the equilibrium approximation and the explicit calculation of $e$ and $i$ has been made using a self consistent giant planet formation model. Moreover, out of equilibrium effects can be important, not only at the beginning of the formation of a planet.  When planets migrate they can enter regions where planetesimals are in principle cold (low values of $e$ and $i$), or 
already excited if there is another planet growing in the neighbourhood. Depending on the ratio between the stirring and the migration timescale, one can expect some cases where the 
equilibrium approximation may not be  accurate.

We study  the formation of giant planets considering a self-consistent model for the interplay between the disc evolution, accretion by the growing planets and gas driven migration. In this paper, we focus our
study on the formation of single planets,  looking in detail at the dependence of planetary growth upon the planetesimal size and the 
differences in the final results between the equilibrium approximation and the explicit calculation of $e$ and $i$, all together with planetary migration. We do this analysis in four steps. First we consider the formation of 1 M$_\oplus$ planet, 
neglecting the presence of an envelope, of planetary migration and of disc evolution (Sect. \ref{Sect:1ME}). Second, we compute the full formation of a planet (except migration), which is assumed to be over when the gaseous
 component of the disc disappears. We keep the in situ formation hypothesis to ease the analysis (Sect. \ref{Sect:full_wo_mig}). Third, we allow for gas driven migration during the 
 formation of the planet (Sect. \ref{Sect:single_full_formation}). Fourth,  we generalise the examples presented in the third step by considering a wide range of plausible protoplanetary discs and initial 
 locations for the embryo to have an overview of all possible outcomes (Sect. \ref{Sect:population}). 
\begin{figure*}
\centering
 \includegraphics[width= 0.7\textwidth, angle= -90]{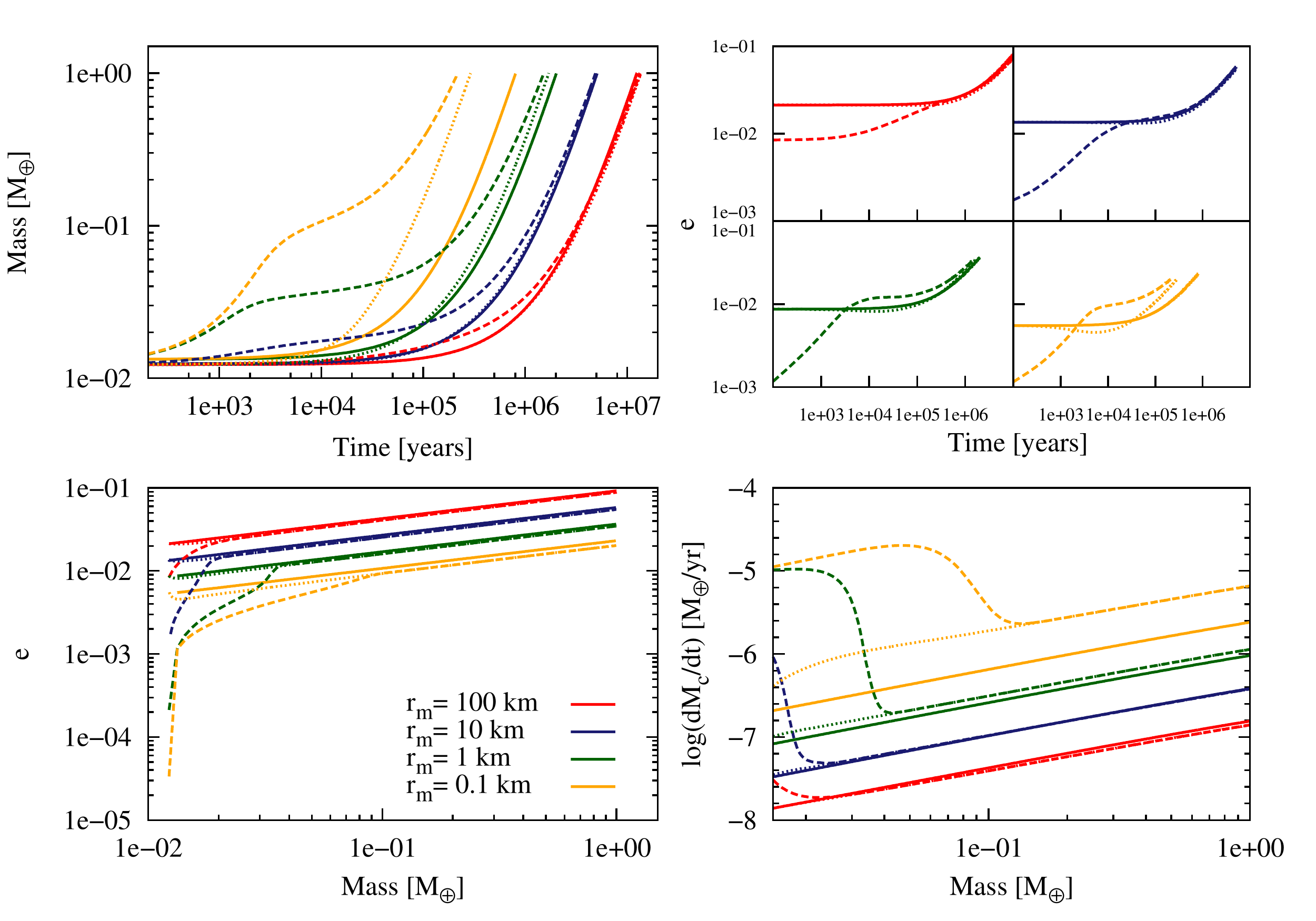} 

   \caption{In situ formation of 1 M$_\oplus$ planet at 6 AU for different radii of accreted planetesimals (red, 100 km; blue, 10 km; green, 1 km; orange,  0.1 km). Solid line is for the equilibrium approximation of eccentricities and inclinations (see Eqs. (\ref{eq:e_eq}) - (\ref{eq:i_eq})). Dashed and dotted lines are for the explicit calculation of  $e$ and $i$  solving the differential equations  Eqs. (\ref{eq:diff_eqe}) - (\ref{eq:diff_eqi}). Two different initial values, $e_0$, $i_0$, were considered:  dashed line represents the case  where $e_0$ and $i_0$ are given by Eqs. (\ref{eq:i_eqmm}) - (\ref{eq:e_eqmm}), while dotted line assumes $e_0$, $i_0$ to be the initial equilibrium values of the equilibrium case.  Top-left panel depicts the mass growth of the protoplanet as a function of time.  Top-right panel shows the time evolution of the eccentricity.  Bottom-left panel  shows the eccentricity as a function of the mass of the embryo. Clearly, in all cases, after an out of equilibrium state, equilibrium is attained. Bottom-right panel plots the accretion rate of solids as a function of the mass of the embryo. Note that, as a function of the planet mass, the difference in the accretion rate is only a consequence of the difference in $e$ and $i$.     }
    \label{Fig:1ME}
    \end{figure*}

\subsection{Formation of 1 M$_\oplus$ planet}
\label{Sect:1ME}

We analyse first the formation of 1 M$_\oplus$ planet (we stop the calculation when this mass is reached), neglecting the presence of an envelope, planetary migration and disc evolution. 
The point of this case is to focus on the initial stages of the accretion and analyse the importance of the size of the accreted planetesimals and the implications for the growth of the embryo 
when considering  the equilibrium approximation. It is because we want to show clearly the consequences of these assumptions that we neglect other physical processes that act simultaneously. 
This means, for example, that for a fixed mass of the embryo and independently of the elapsed time, the state of the protoplanetary disc is the same in terms of surface density (solids and gas). 
The same applies for the capture radius of the planet.  In this example, the embryo is assumed to be located at 6 AU, where the initial solids surface density is $\Sigma_m= 10 \, \mathrm{g\,cm^{-2}}$ 
and the density of the nebular gas is $\rho_\mathrm{gas}= 2.4\, \times 10^{-9} \mathrm{g\,cm^{-3}}$. For this disc the snow line is at 3.5 AU. The initial surface density profile of the disc is given by Eq. (\ref{eq:init_sigma}) 
with $\gamma=0.9$ and $a_\mathrm{C}= 127$ AU. The initial mass of the embryo is 0.01  M$_\oplus$. For the equilibrium approximation we adopt Eqs. (\ref{eq:e_eq}) - (\ref{eq:i_eq}) to calculate 
the values of $e$ and $i$. For the explicit calculation of  the eccentricity and inclination of planetesimals we solve Eqs. (\ref{eq:diff_eqe}) - (\ref{eq:diff_eqi}). To solve Eqs. (\ref{eq:diff_eqe}) - (\ref{eq:diff_eqi}), 
initial conditions for $e$ and $i$ must be given.  We  consider two possibilities for the initial conditions that we think bracket  the parameter space. On one hand, we consider 
that the planetesimal disc is initially cold, and planetesimals' eccentricities and inclinations are given by the equilibrium value between their mutual stirring and the gas drag. These values can be 
derived by equating the stirring timescale and the damping timescale, that results in
\begin{eqnarray}
e_\mathrm{eq}^{m-m}&=&2.31\frac{ m^{4/15}\Sigma^{1/5} a^{1/5}\rho_m^{2/15}}{C_\mathrm{D}^{1/5}\rho_\mathrm{gas}^{1/5}M_{\star}^{2/5}},\label{eq:e_eqmm}\\
i_\mathrm{eq}^{m-m}&=&\frac{1}{2}e_\mathrm{eq}^{m-m}.
\label{eq:i_eqmm}
\end{eqnarray}
This means that we are assuming that the embryo instantaneously appears in an unperturbed planetesimal disc. The other extreme situation is to assume a hot disc, where planetesimals are already 
excited by the embryo and their initial eccentricities and inclinations are those corresponding to the value of equilibrium between the stirring of the embryo (0.01  M$_\oplus$) and the gas drag, approximated by Eqs. (\ref{eq:e_eq}) - (\ref{eq:i_eq}). 
The initial values of $e$ for these cases are given in table \ref{table:initial_e}, where $e_\mathrm{eq,0}^{m-m}$ correspond to Eq. (\ref{eq:e_eqmm})  and $e_\mathrm{eq,0}$ to Eq. (\ref{eq:e_eq}). The initial values 
of $i$ are assumed to be $e/2$ ($i_0=e_0/2$). We perform calculations for the equilibrium approximation and the two sets of initial conditions for the explicit calculation, for four radii of  the accreted planetesimals: 100, 10, 1 and 0.1 km. 
\begin{table}
\caption{Two sets of initial values of the eccentricity for the explicit calculation of its evolution. }
\label{table:initial_e}
\centering                       
\begin{tabular}{c c c }        
\hline
         $r_m$      &   $e_\mathrm{eq,0}^{m-m}$ & $e_\mathrm{eq,0}$\\
 \hline
  	100 km & 8.4 $\times 10^{-3}$ & 2.13 $\times 10^{-2}$ \\
	10   km & 1.3 $\times 10^{-3}$ & 1.35 $\times 10^{-2}$ \\
	1     km &  2.0 $\times 10^{-4}$ & 8.70 $\times 10^{-3}$ \\
	0.1 km &  3.3 $\times 10^{-5}$ & 5.53 $\times 10^{-3}$ \\
\hline
\end{tabular}
\end{table}

Fig. \ref{Fig:1ME} shows the results of the simulations mentioned above. The top-left panel depicts the mass growth of the planets as a function of time. The first evident thing from this plot is 
the timescale difference in the formation of 1 M$_\oplus$ planet depending on the size of the accreted planetesimals: while for accreted planetesimals of 100 km it takes $\sim 10^7$ yr to grow 
from 0.01 M$_\oplus$ mass to 1 M$_\oplus$, it takes $\sim 10^5$ yr in the case of 0.1 km planetesimals. This difference in the growth timescale is entirely due to the fact that large planetesimals are less damped by gas drag than the smaller ones. Hence, while they are stirred up by the massive embryos to about the same value, they keep larger $e$ and $i$ values (bottom-left panel) making the accretion process much slower.  Keep in mind that the disc does not evolve and planets do not migrate, so differences depend  only on the planetesimal size. Note that for a fixed mass of the embryo the accretion rate of solids differs in two orders of magnitude 
between the two extreme cases (bottom-right panel). 

If we now turn our attention to the differences in growth rate for a fixed planetesimal size, but for different approaches in the calculation of the eccentricities and inclinations, we see that adopting 
the equilibrium approximation may lead to significant differences  in the mass of the planet, more evident when the accreted planetesimals are small (top-left panel, compare solid line with dashed
or dotted lines of the same colour). For 100 km and 10 km planetesimals we do not see deviations from equilibrium when the initial conditions for $e$ and $i$ are given by Eqs. (\ref{eq:e_eq}) - (\ref{eq:i_eq}) 
(red and blue dotted lines, top-right panel). If the initial values for $e$ and $i$ are given by Eqs. (\ref{eq:e_eqmm}) - (\ref{eq:i_eqmm}) the equilibrium values given by Eqs. (\ref{eq:e_eq}) - (\ref{eq:i_eq}) 
are reached in an almost negligible fraction of the formation timescale.  We conclude that the equilibrium approximation for $e$ and $i$ gives results that nicely agree with more complex calculations, 
as far as the planetesimal size is relatively big or the evolution time is sufficiently long. However, as shown on the same picture, the growth of a small planet is very long when considering such massive planetesimals, and the formation
of a gas giant under these conditions is highly compromised.

\begin{figure}[]
\centering
\includegraphics[width=0.35\textwidth, angle = -90]{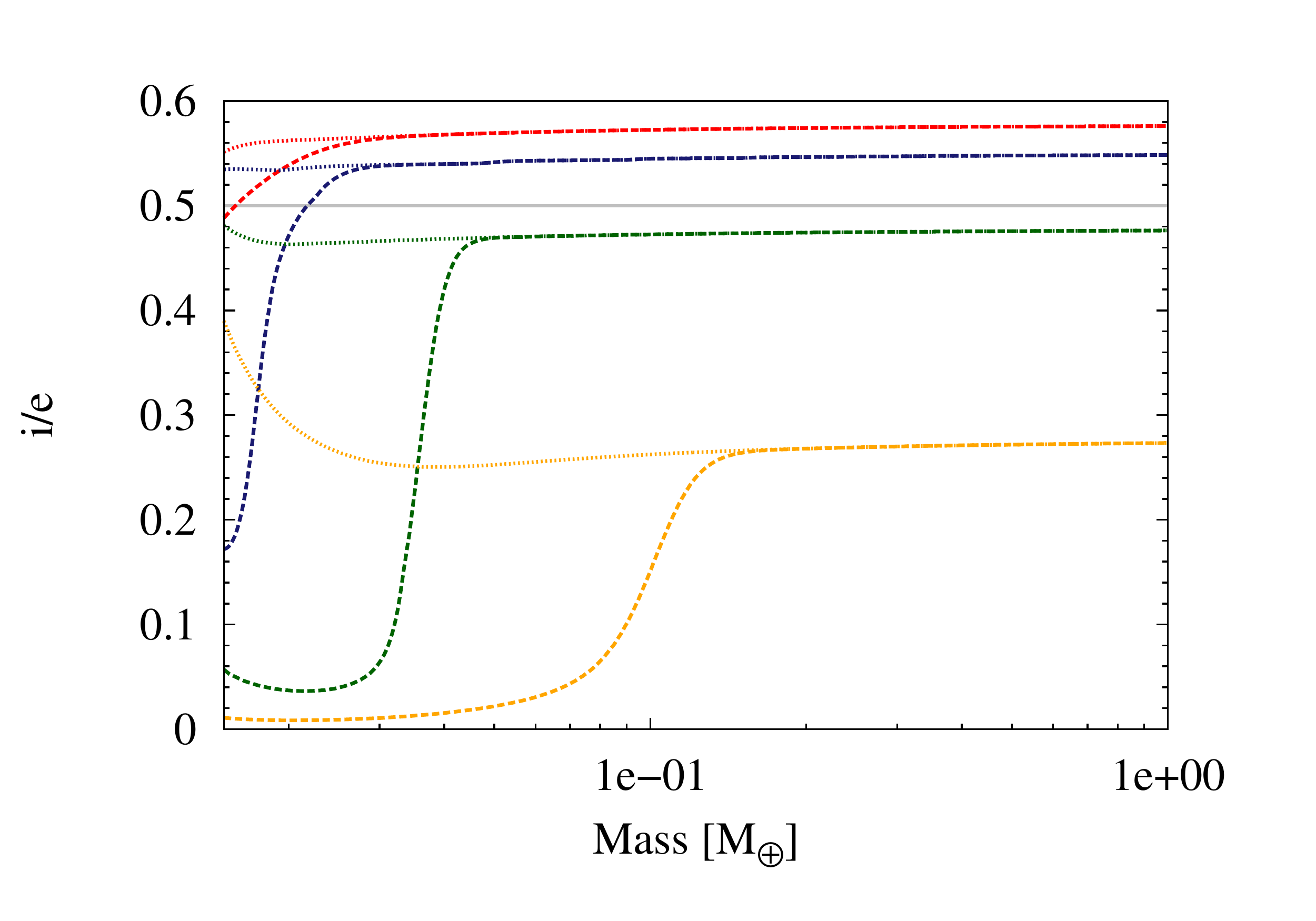}
 
   \caption{Inclination-eccentricity ratio for the out of equilibrium cases of Fig. \ref{Fig:1ME}.  Red is for 100 km planetesimals, blue for 10 km, green for 1 km and orange for 0.1 km.  Dotted line represents the case of an initially hot disc while dashed line of an initially cold one. The grey line is the standard equilibrium value, $\beta_\mathrm{eq}=i/e=1/2$.  }
              \label{Fig:beta_m}
    \end{figure}

The situation becomes more critical when we consider smaller planetesimals. Deviations from equilibrium are evident regardless the initial values adopted for $e$ and $i$, especially for $r_m=0.1$ km. 
Nevertheless, equilibrium is always attained, although the equilibrium values for $e$ and $i$ are lower than those given by  Eqs. (\ref{eq:e_eq}) - (\ref{eq:i_eq}), especially in the case of  the inclination. 
As we can see from Fig. \ref{Fig:beta_m}, the approximation  $\beta_\mathrm{eq}=i/e=1/2$ is not very good for smaller planetesimals. Note that the real equilibrium value of $\beta$ depends upon the 
planetesimal size. While $\beta_\mathrm{eq}=1/2$ is a good approximation for planetesimals larger than $\sim$ 1 km, it is not the case for smaller planetesimals, which tend to have lower inclination values 
than $1/2\,e$. We find that for  small planetesimals, the velocity regime is in the limit between the high and medium velocity regime ($\tilde{e}$ , $\tilde{i}$ $\lesssim$ 2), therefore eccentricities are more 
effectively excited than inclinations (see Ohtsuki et al. 2002). The low values of $e$, $i$ and $\beta$  increase the accretion rate, as can be seen in the bottom-right panel of Fig. \ref{Fig:1ME}, speeding up the formation of the embryo 
(relative to the equilibrium approximation case) . 

\subsection{Planet formation without migration}
\label{Sect:full_wo_mig}
   \begin{figure*}
   \centering
   \includegraphics[width= 0.7\textwidth, angle = -90]{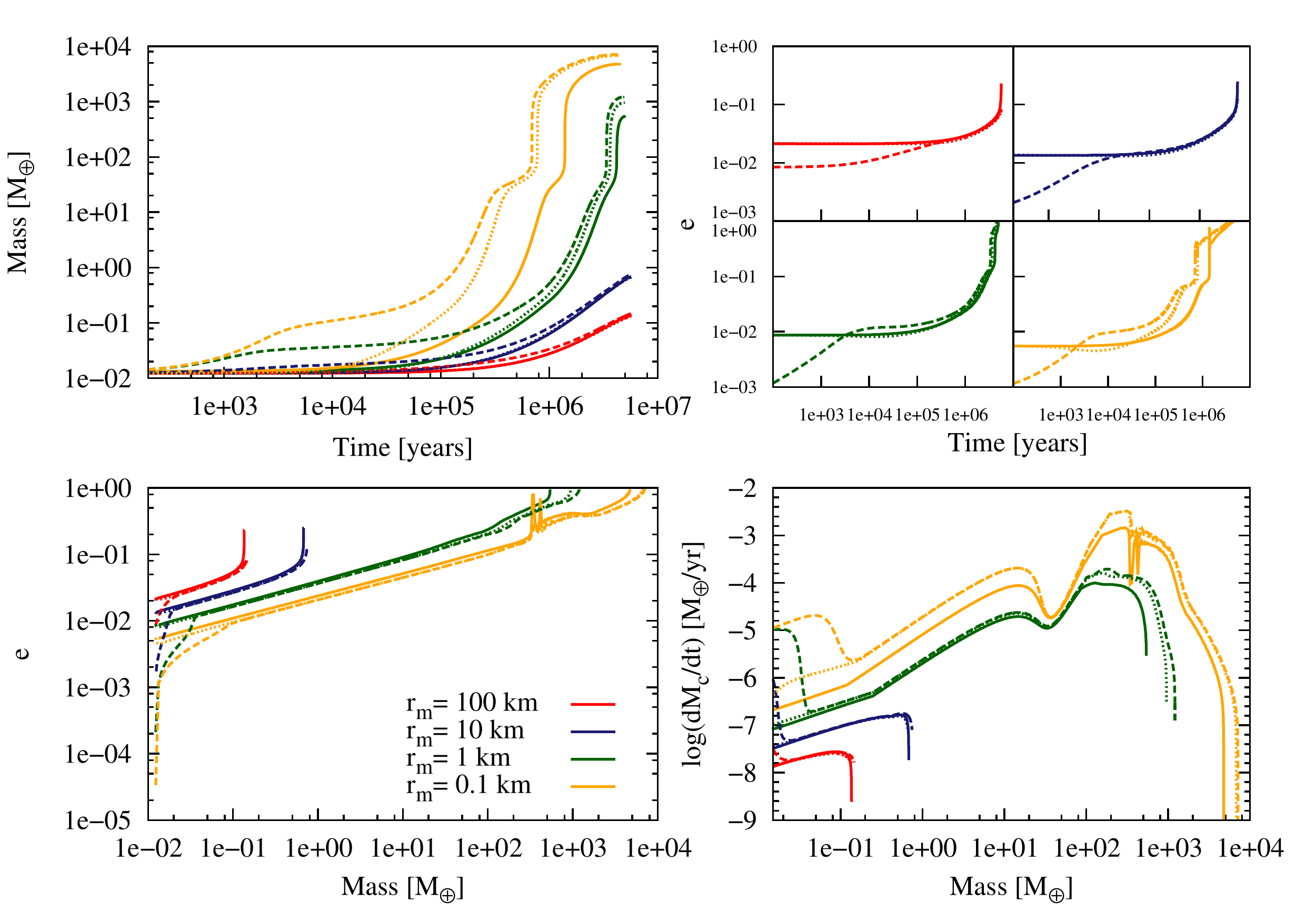} 

   \caption{In situ planetary formation at 6 AU for different radii of the accreted planetesimals (red, 100 km; blue, 10 km; green, 1 km; orange,  0.1 km). The same line code as in Fig. \ref{Fig:1ME} is adopted. The formation of the planet ends when the gas component of the disc is dispersed. }
              \label{Fig:insitu_full}
    \end{figure*}

In this section we analyse the complete formation of a planet for the same cases as before. Now,  the planet accretes mass (solids and gas) as long as the disc is still present.  
This means that we allow for disc evolution (Sect. \ref{disc}), therefore planet formation is considered to be over when the disc disperses. The presence of the gaseous envelope 
is taken into account for the calculation of the capture radius. However,  we still neglect the migration of the embryo to ease the analysis of the dependence on planetesimals' $e$ and $i$. 

As it has been already shown in other works, the presence of the envelope increases the capture radius, speeding up the formation of the planet.  The enhancement in the capture 
cross section makes the accretion of solids more effective. This effect is more noticeable for small planetesimals. Compared to the cases of Sect. (\ref{Sect:1ME}), the formation timescale of a 
1 M$_\oplus$ can be reduced up to  $\sim$ 35 \% (for the smallest planetesimals, $r_m=0.1$ km), due only to the enhancement in the capture cross section.  The presence of an atmosphere, 
even if its mass is negligible compared to the total mass of the planet, has to be considered for embryos as small as 0.1 M$_\oplus$ as it plays an important role for the accretion of solids.

Fig. \ref{Fig:insitu_full} is the analog of Fig. \ref{Fig:1ME} for the complete formation of the planets. As the disc disperses after 6 Myr, in the cases where the accretion of planetesimals is slow 
($r_m= 100, 10$ km), the final masses of the planets are lower than 1 M$_\oplus$. The differences in growth observed in the previous section as a function of the planetesimal size have dramatical 
consequences for the final mass of the planet, which can be $\sim$ 0.1 M$_\oplus$ if the accreted planetesimals have a radius of 100 km,  $\sim$ 0.8 M$_\oplus$ for planetesimals of 10 km, 
 $\sim$ 1200 M$_\oplus$ (3.7 Jupiter masses, M$_\mathrm{J}$) for 1 km planetesimals, and 7100 M$_\oplus$ (22 M$_\mathrm{J}$) for 0.1 km planetesimals. These numbers evidence how
 differences in the accretion rate of solids (here regulated by the size of the accreted planetesimals, bottom-right panel)  impact on the final mass of a planet and the non-linear aspect of planet formation
 in the core accretion model: once the critical mass is attained, the very rapid accretion of gas leads rapidly to massive planets. The fact that bigger planets form when the accreted planetesimals 
 are small is a consequence of the gas drag, which operates in two ways that combine positively:  nebular gas drag is more effective in damping planetesimal's eccentricities and inclinations when 
 planetesimals are small (bottom-left panel) and atmospheric gas drag is able to deflect more distant planetesimals' trajectories  therefore enlarging the capture radius of the planet. In fact, for the cases of 
 1 km and 0.1 km accreted planetesimals, embryos grow to become big giants. When a massive solid embryo is formed it triggers the accretion of gas, leading to the formation of a gas giant planet
  (green and orange lines).

 Indeed, planets can end up very massive if they enter the runaway phase of gas. As explained before, during the attached phase, the accretion of gas is a result of solving the equations presented in Sect. \ref{equations}. The planet will remain attached to the disc until its accretion rate exceeds the maximum amount of gas that the disc can deliver. When this condition is not any more satisfied, it goes into the detached phase. During the detached phase, the maximum accretion rate is given by Eq. (\ref{maxdotgas}).  As in M09 and M12, here we follow the results of  Kley \& Dirksen (2006) for the accretion of gas during the detached phase. Planet-disc interactions can lead to eccentric instabilities, which means that the planet can enter regions that are outside its gap and have full access to the gas present in the disc, with no limitation for accretion except for the ability of the disc to supply it. Although it is not clear whether all planets can suffer from an eccentric instability, for the sake of simplicity we assume this is the general situation in our simulations. On the other hand, other works (e.g. Lissauer at al. 2009) include a limitation for the accretion of gas when the planet opens a gap in the disc, as they consider that planets are in circular orbits. Comparatively, this assumption leads to less massive planets. 

   \begin{figure*}
   \centering
   \includegraphics[width= 0.5\textwidth, angle = -90]{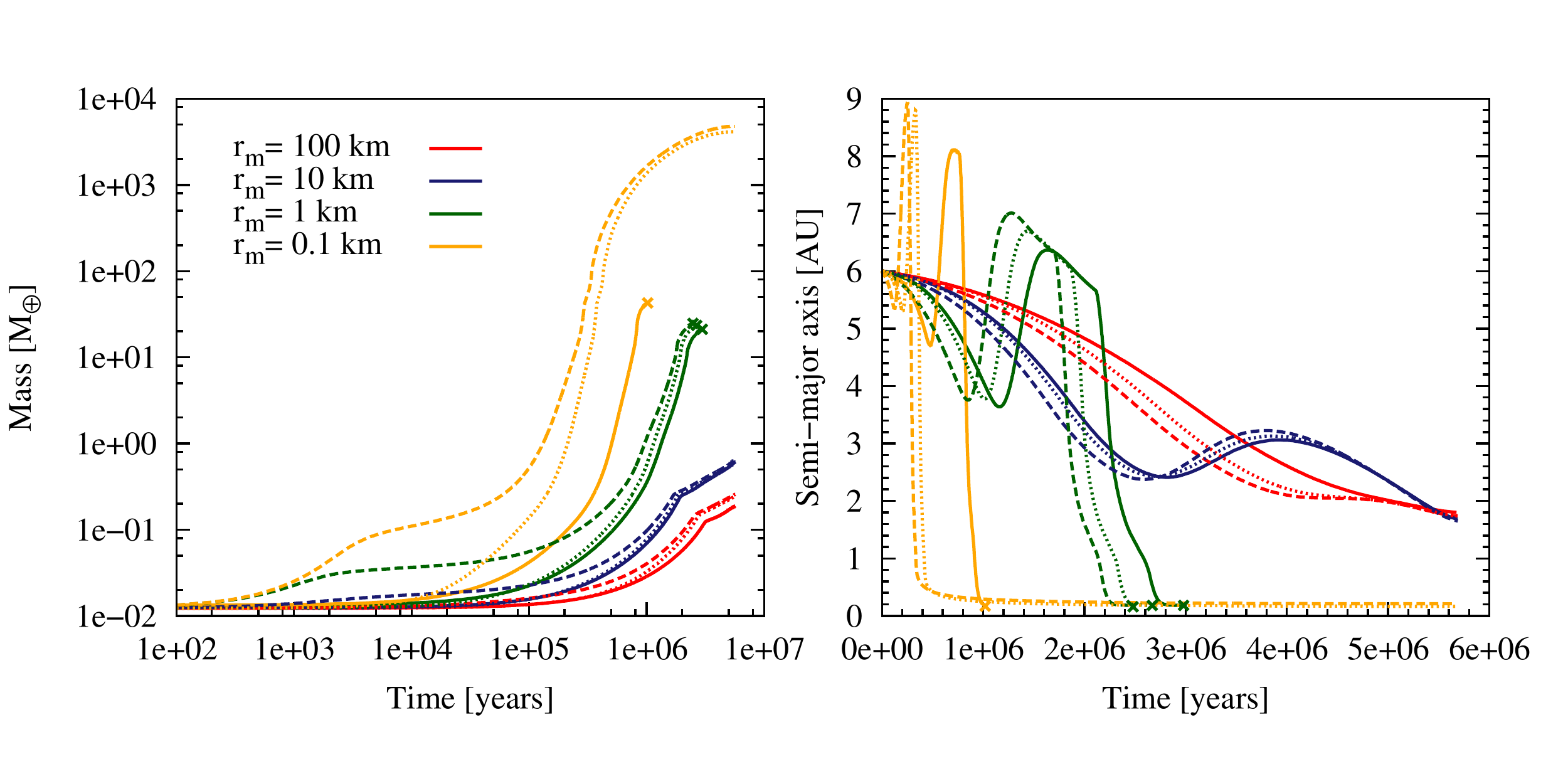} 

  \caption{Full planetary formation, now allowing for the migration of the embryos, for different radii of the accreted planetesimals (red, 100 km; blue, 10 km; green, 1 km; orange,  0.1 km). The same line code as in Fig. \ref{Fig:1ME} is adopted. Embryos are initially located at 6 AU. Left panel shows the cumulative mass of the embryo (solids plus gas) as a function of time (in logarithmic scale).  Right panel depicts the migration path of the protoplanet (in linear scale). Note that for the accretion of 1 km  planetesimals,  the embryos are lost in the Sun (indicated with crosses in the plot). The same happens for the planet that grows by accretion of 0.1 km planetesimals, in the equilibrium approximation (orange solid line). When planets are lost, calculations are stopped. }
                \label{Fig:complete}
    \end{figure*}

The equilibrium approximation and the explicit calculation of $e$ and $i$ also bring differences in the final mass of the planet. Whether for large planetesimals (100 and 10 km) the equilibrium 
approximation works fine as showed in the previous section, for smaller planetesimals it might not be the case.  For $r_m=1$ km,  the final mass of the planet in the equilibrium approximation is 
537 M$_\oplus$,  whether with the explicit calculation considering an initial hot disc is 950 M$_\oplus$ and with an initial cold disc, 1200 M$_\oplus$. For $r_m=0.1$ km,  the final mass of the planet 
in the equilibrium approximation is  4745 M$_\oplus$,  whether with the explicit calculation considering an initial hot disc 6745 M$_\oplus$  and with an initial cold disc, 7100 M$_\oplus$.  Differences 
in mass are the consequence of a timing effect.  When the planet grows faster, the damping produced by the nebular gas is more effective because the gas density in a younger disc is larger 
than in an older one. Then, the faster an embryo grows, the more it profits from the nebular gas drag. Moreover, the cross-over mass (the mass of the core for which the mass of the envelope 
equals the mass of the core), which in this case is $\sim$ 30 M$_\oplus$,  is achieved earlier when the accretion rate of solids is bigger.  When a protoplanet reaches the cross-over mass its 
growth is dominated by the accretion of gas, that is already in the runaway regime. If the protoplanet starts the accretion of gas when the disc is younger and more massive, it  provides a larger reservoir of gas
 which translates, in the end,  in a bigger planet.

It is important to remark that Eqs. (\ref{eq:e_eq}) - (\ref{eq:i_eq}) are obtained assuming that the stirring timescale is equal to the damping timescale (Thommes et al. 2003), and the  equilibrium 
approximation  ($i=0.5\, e$).  On the other hand, Eqs. (\ref{eq:diff_eqe}) - (\ref{eq:diff_eqi}) explicitly solve the coupled evolution of $e$ and $i$ given a set of initial conditions. This means that out 
of equilibrium states are allowed (e.g. as seen at the beginning of the calculations) and equilibrium states are reached naturally and without a fixed ratio between $e$ and $i$. It is clear that the 
explicit resolution of the differential equations is a better physical approach. We have included  here calculations with the equilibrium approximation just for comparison. Results show that the 
equilibrium approximation works fine for larger planetesimals, but overestimates the excitation of smaller ones making their accretion less effective. This delays the whole process of planet formation. 
As the planet is embedded in a disc with a finite (and short) lifetime, this delay impacts in the final mass of the planet. When solving explicitly the equations of $e$ and $i$, initial conditions for these 
quantities are needed. This is a problem because we can not be certain about the state of the disc at the beginning of our calculations. So assumptions for the initial values can not be avoided. 
The ``initially cold disc''  favours the growth of a planet allowing for high accretion rates in the first thousands of years (Fig.\ref{Fig:insitu_full}, bottom-right panel). In the ``initially hot disc'' this 
effect is not present as departures from equilibrium are smaller and therefore equilibrium is attained faster (keep in mind that the equilibrium attained when solving the differential equations can be 
different to the equilibrium approximation for the case of small planetesimals). However, final results do not strongly depend upon the initial choices.  The differences in the final mass are of $20\%$ and 
$5\%$ for $r_m= 1$ km and $r_m= 0.1$ km respectively. Given all the uncertainties of the model these differences are acceptable. 

\subsection{Planet formation with migration}
\label{Sect:single_full_formation}

We now complete our analysis including the migration of the protoplanets. The migration model is the one presented in Dittkrist et al. (in prep.) and Mordasini et al. 2010. Fig. \ref{Fig:complete} shows the  total mass of the planet and its semi-major axis as a function of time for the same cases as in the previous sections.  Clearly, the situation is very different from the 
in situ hypothesis. To start with, in all the cases the final location of the planet is far from its initial location. Actually, the three calculations for 1 km planetesimals result in lost planets (planets cross the 
inner edge of the disc and are considered to fall into the central star). The same is the case for the equilibrium approximation of  0.1 km planetesimals.

\begin{figure}
\center

\includegraphics[width=0.35\textwidth, angle = -90]{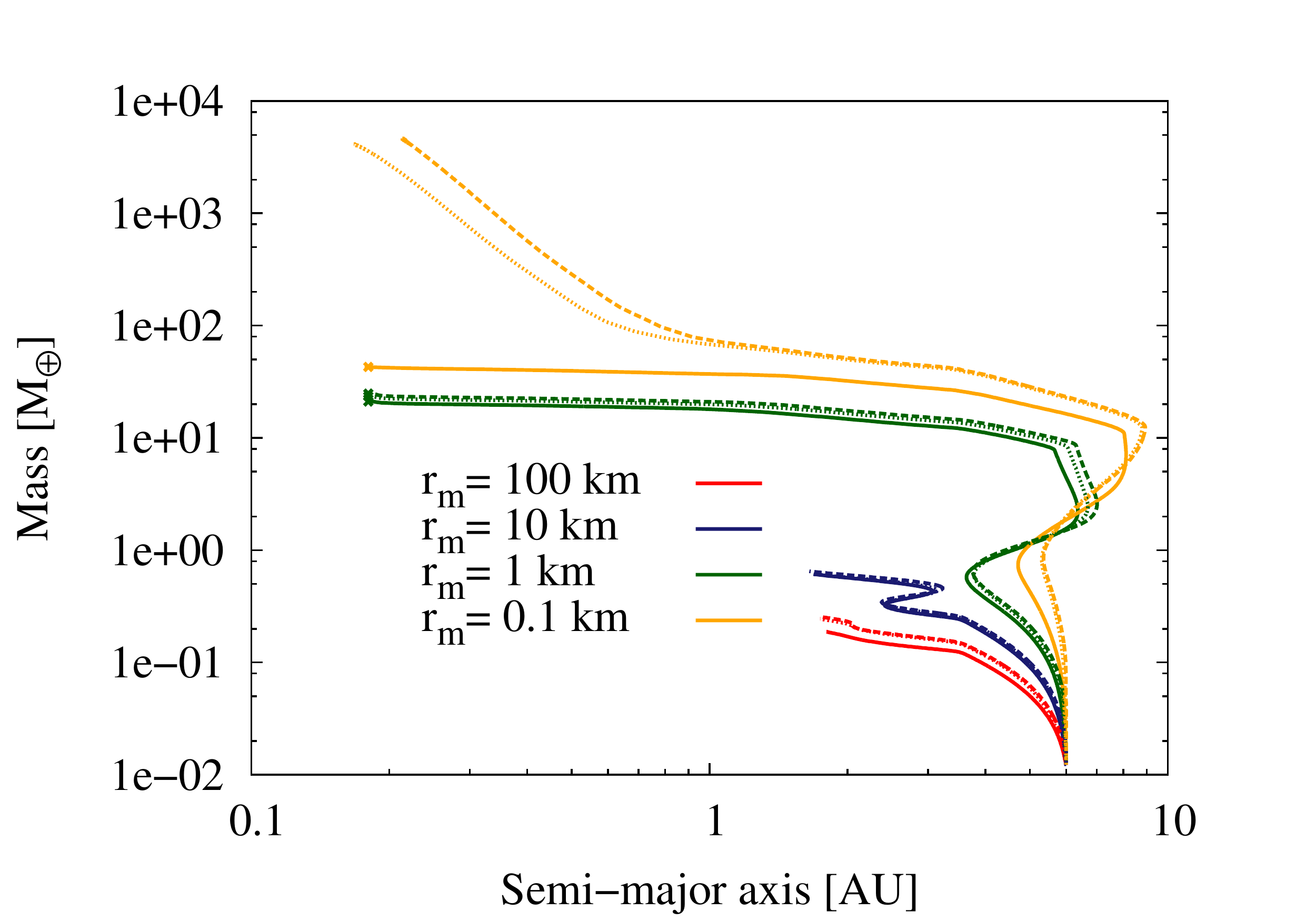}

   \caption{Mass versus  semi-major axis for the cases shown in Fig. \ref{Fig:complete}. Note that the accretion rate of solids plays a major role not only in the growth of the planet but also in its migration path.  }
             \label{Fig:complete_m_a}
    \end{figure}

For the cases where planets are not lost into the star, we find that the final masses for accreted planetesimals of $r_m=$ 100 km and 10 km are independent of the way $e$ and $i$ were computed 
($\sim$ 0.2 M$_\oplus$ and 0.7 M$_\oplus$ respectively). The final locations are also similar for the different considerations for  $e$ and $i$ ($\sim$ 1.7 AU for both $r_m=$ 100 km and 10 km).  
For a fixed planetesimal size, the migration paths for the different assumptions on the eccentricities and inclinations are somehow shifted in time but are similar if we consider them as a function 
of mass (Fig. \ref{Fig:complete_m_a}).

In the case of $r_m= $1 km, protoplanets are lost in the central star when their mass  is $\sim$ 20 M$_\oplus$, most of which is in the solid core. At $\sim$ 8 $M_\oplus$ the protoplanet undergoes inward migration in the adiabatic saturated 
regime\footnote{The regime is called ``adiabatic'' when the local cooling timescale in the disc is larger than the time it takes for a parcel of gas to make a U-turn on the horse-shoe orbit close to the planet in the corotation region. A regime is called ``saturated''  when the contribution of the corotation region to the angular momentum exchange is reduced by the the ratio of viscous timescale and libration timescale. In this case, the migration behaviour will get dominated by the angular momentum exchange  at the Lindblad resonances.}  
(see Paardekooper et al. 2010, 2011, Mordasini et al. 2010, Dittkrist et al.). This timescale turns out to be shorter than the accretion timescale. The protoplanet  covers 
$\Delta a \simeq$ 5 AU in  $\sim 7 \times 10^5$ yr. In this time it doubles its mass which, however,  is not big enough for the planet to enter type II migration. Planet migration in general 
slows down when the protoplanet is able to open a gap in the disc (type II migration) which, as a rule of thumb, happens when the planet mass is $\sim 100$  M$_\oplus$. This situation is not reached 
in this case, where accretion is too slow compared to migration, resulting in the loss of the forming planet. For $r_m=$ 0.1 km  the differences between the equilibrium approximation 
and the explicit calculation of $e$ and $i$ are more dramatic: adopting the equilibrium approximation leads to the loss of the planet in the central star (for the same reason as in the previous case, 
the growth rate is very slow) whether for the explicit calculation of $e$ and $i$, although the planet ends in an orbit very close to the central star ($\sim 0.2$ AU), the previous growth of the embryo is 
fast enough to enable a large accretion of gas. In this case the planet reaches a mass that enables it to switch to type II migration. The planet decelerates its migration speed until it stops. The final masses are: 13 M$_\mathrm{J}$ for in initially hot disc and 15 M$_\mathrm{J}$ for an initially cold disc.
The fate of an embryo (becoming a big or a low mass planet, surviving or being lost in the central star), as it is shown here, depends upon the size of the accreted planetesimals 
and on the assumptions we adopt to describe their dynamics, as these strongly impact on the accretion rate of solids and therefore on the whole formation process through the regulation
of the growth timescale.  Here we have shown the interplay 
between the evolution of the protoplanetary disc, the growth of the protoplanet and the operation of migration.  Note that in all the cases we are considering the same disc, that globally evolves with 
time in the same way. Differences in the local evolution of the disc arise due to accretion (solids and gas accreted by the planet are removed from the disc) and ejection of planetesimals (when the 
planet is massive enough). The fact that in Fig. \ref{Fig:complete_m_a}, for the same mass of a protoplanet, the location in the disc can be different is a consequence of this interplay: 
protoplanets reach a certain location earlier or later in the evolution of the disc, depending on their growth rate. The state of the disc and the mass of the protoplanet at that moment determines its 
migration rate. So, independently of what regulates the growth of the planets, the examples here show that planet's growth and migration rate are tightly coupled.  

\begin{figure*}
   \centering
   \includegraphics[width= 0.7\textwidth, angle = -90]{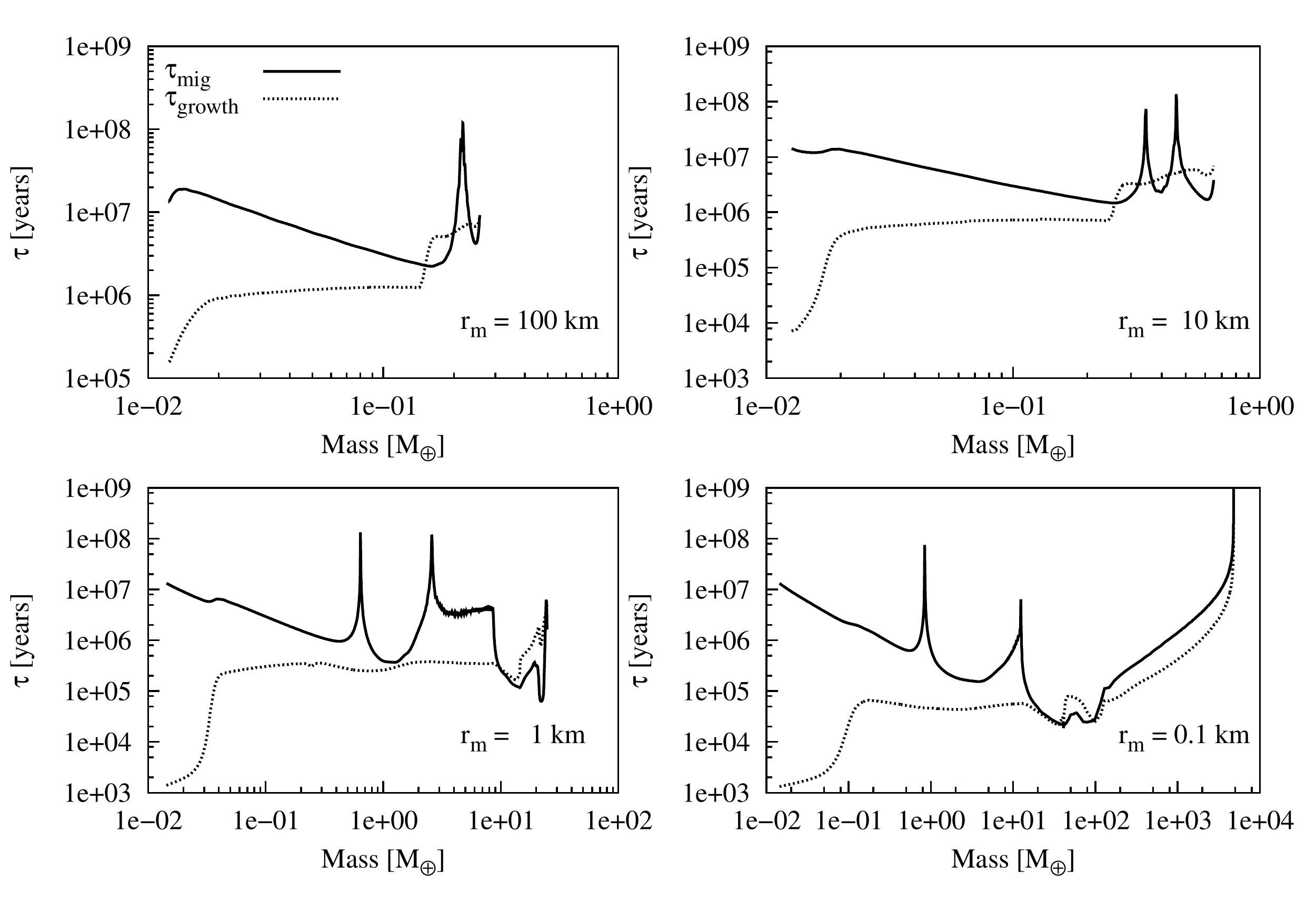} 

  \caption{Protoplanet's growth and migration timescales ($\tau_{\mathrm{growth}}=M/|\dot{M}|$ and $\tau_{\mathrm{mig}}=a/|\dot{a}|$ respectively) as a function of the protoplanet's mass for the four different sizes of the accreted planetesimals of Fig. \ref{Fig:complete} (corresponding to the dashed line in Fig.(\ref{Fig:complete})).  The peaks in the migration timescale correspond to the changes in the sense of migration of the protoplanet. From the two bottom panels, it can be seen that when the protoplanet's mass is $\gtrsim 10 \, \mathrm{M_\oplus}$ both timescales become comparable and eventually the migration timescale becomes shorter than the accretion timescale, leading to a fast migration of the protoplanet with very little accretion. In the case of 1 km planetesimals the protoplanet growth is too slow  to gain enough mass to change the type of migration before it is lost in the star (when its mass is $\sim 30 \, \mathrm{M_\oplus}$). In the case of 0.1 km planetesimals, the protoplanet can grow more massive because its accretion rate is larger than in the previous case, then it is able to  open a gap in the disc and migration switches to type II, preventing it to fall in the star. }
             \label{Fig:timescales}
    \end{figure*}

Fig. \ref{Fig:timescales} shows a comparison between the migration timescale (\mbox{$\tau_\mathrm{mig}=a/|\dot{a}|$}) and the protoplanet's growth timescale (\mbox{$\tau_{\mathrm{growth}}=M/|\dot{M}|$}) for the four sizes 
of the accreted planetesimals we have considered, solving explicitly the equations for $e$ and $i$ and adopting an initially cold planetesimal disc.  For accreted planetesimals of 100 km and 10 km, 
 the jump in the growth rate of the planets at  $M \gtrsim 0.1$  M$_\oplus$ corresponds to the crossing of the ice line. Planetesimals in the inner region (inside the ice line) are denser and the gas drag is less effective on them 
(see Eqs. (\ref{eq:drag_quadratice}) - (\ref{eq:drag_quadratici}) and (\ref{eq:tau_drag})), therefore their random velocities are higher and accretion rates are lower. Also, the solids surface density is lower, which contributes to decrease the accretion rate.  
The peaks in the migration timescale correspond to changes in the sense of migration\footnote{Depending on the regime -  isothermal \textit{versus} adiabatic, and saturated \textit{versus} unsaturated,
the migration can be inward or outward, see Paardekooper et al. 2010, 2011, Mordasini et al. 2010, Dittkrist et al., in prep.}.

When the accreted planetesimals have a radius of 1 km,  the migration timescale becomes lower than the growth timescale when the mass of the protoplanet is $\sim 10\, \mathrm{M}_\oplus$. Planetesimals relative velocities are very high in the neighbourhood of the protoplanet and accretion becomes more difficult as the protoplanet increases its mass. The protoplanet grows slowly and migrates fast. This situation is never reverted and the protoplanet is lost in the central star. Being the migration rate high and the accretion rate not enough to counteract it,  the protoplanet migrates without limit, almost at a constant mass, until it reaches the central star. 

In the case of accreted planetesimals of 0.1 km, the protoplanet's migration timescale is shorter than the protoplanet's accretion timescale (just as in the former case) for the ``critical'' mass interval of a few tens  to around one hundred of Earth masses. However, in this case accretion proceeds fast enough for the protoplanet to start runaway accretion of gas before reaching the inner edge of the disc. Therefore the planet is able to accrete mass very fast (gas accretion dominates the growth of the planet), which means that it becomes massive enough to enter type II migration. The migration rate (in type II with gap opening)  decreases as the protoplanet grows in mass. Therefore, being in the runaway phase is what saves the planet from falling into the star. 

The situation can be summarised as follows:  when the growth of the planet is dominated by the accretion of solids in the oligarchic regime (before gas runaway accretion), the growth timescale is proportional to 
$M^{1/3}$ (Ida \& Makino 1993). However, for planets massive enough to trigger rapid gas accretion, the growth timescale is much shorter. On the other hand,  migration typical time scales with 
$M^{-1}$ in type I, and is independent of the planet's mass in type II, being generally much longer than in type I. As a consequence, if a planet migrates in type I and is dominated by the oligarchic growth, it is 
very likely to be lost in the star. It is only if the planet succeeds to become massive enough to start the runaway gas accretion, and quasi-simultaneously enter type II migration, that it can brake before reaching the central star.  Therefore, there is a  critical mass range between $\sim 10 -  100$ M$_\oplus$: a  planet in this mass range is likely to be undergoing inward type I migration and a decreasing growth rate, 
which in this stage is dominated by the accretion of solids. Indeed, being massive, the protoplanet excites the random velocities of planetesimals making its accretion difficult. Therefore the growth of the planet 
is very slow at this stage, as planetesimals can not be effectively accreted. Although this is the pathway to the runaway accretion of gas,  the transition between being dominated by the accretion of solids and the 
accretion of gas, even fast,  is not immediate. On the other hand, inward type I migration in this mass range is still very fast.  Therefore, if planets enter the runaway gas phase they would probably grow massive, change to 
type II migration  and avoid a fatal end in the central star. If they remain very small, type I migration is not very harmful.
Finally,  if they grow up to around Neptune mass while the disc is still young but they do not manage to grow fast enough their migration accelerates and they end in the central star.

\subsection{Exploring the parameter space of initial conditions}
\label{Sect:population}

 One may wonder if the examples presented in the previous section are representative of a general case. In order to answer this question we apply our planet formation model  for a  
 variety of protoplanetary discs (including different lifetimes, masses, metallicities, etc.) and initial locations for the embryo. 

 In this framework, we performed sets of  more than 10 000 simulations. Each set  considers a different size for the accreted planetesimals  ($r_m=$ 100, 10, 1 and 0.1 km).  For these simulations we 
 adopt the disc models described in Sec. \ref{evolution}. The initial mass of the embryo is, in all the cases, 0.01 M$_\oplus$. The initial location of the embryo is varied between 0.2 and 30 AU. For an 
 embryo to start at a certain location, we check that the initial mass of solids at its location is greater than the mass of the embryo. We subtract the mass of the embryo from the initial mass of solids of its 
 feeding zone, which means we are assuming instantaneous, in situ formation for it. Planets grow by accreting solids and gas, and can migrate in the disc, according to the model we presented in the 
 previous sections. The evolution of the eccentricity and inclination of planetesimals are calculated  solving the differential equations (Eqs. (\ref{eq:diff_eqe}) - (\ref{eq:diff_eqi})) all throughout the disc 
 for every time step. As initial conditions for $e$ and $i$ we assume that they are given by  Eqs. (\ref{eq:e_eqmm}) - (\ref{eq:i_eqmm}), what we have called before ``an initially cold disc''. 
 
\begin{figure}
\centering
 \includegraphics[width= 0.35\textwidth, angle = -90]{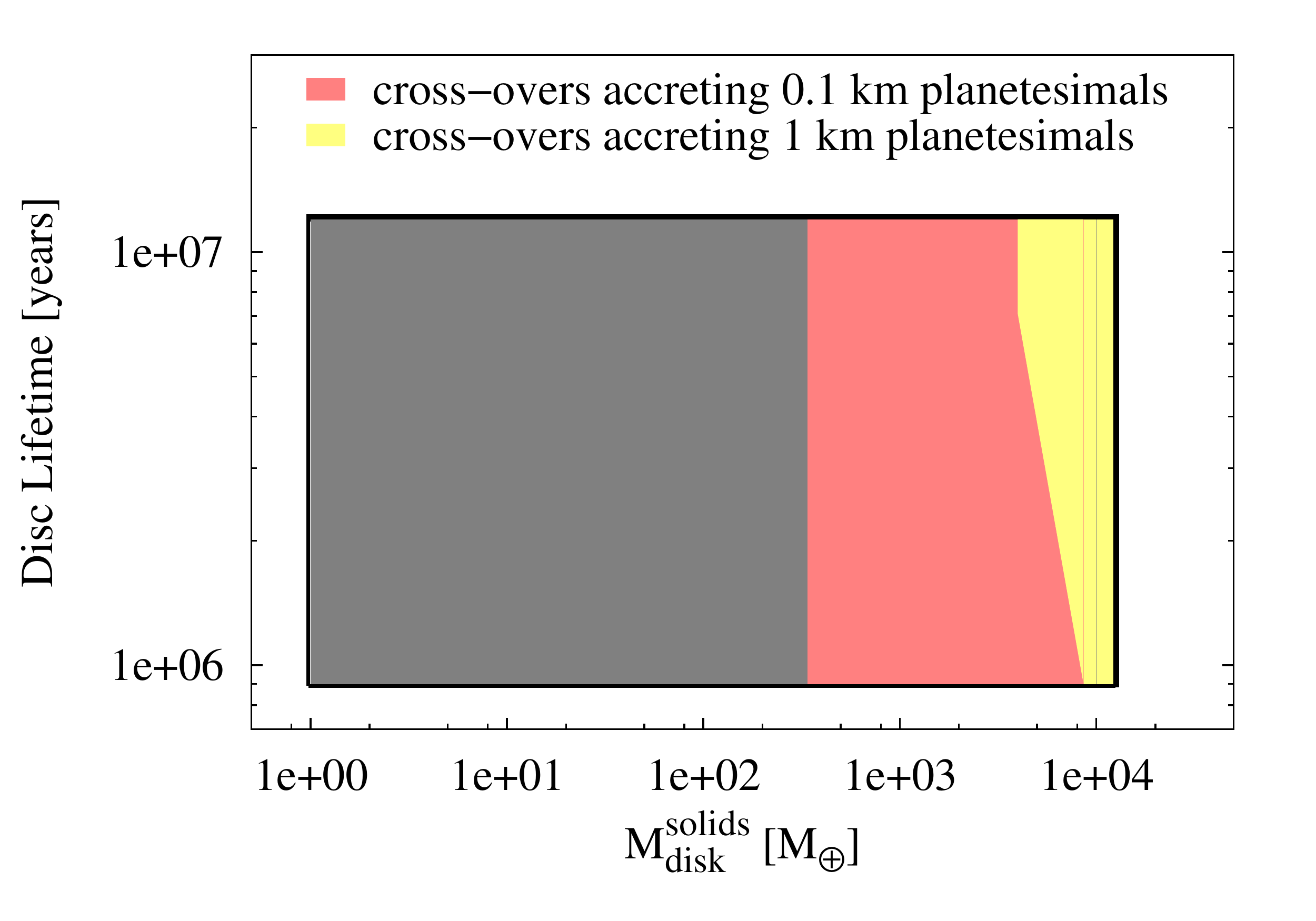} 

   \caption{Lifetime vs. initial solid mass of the protoplanetary discs considered in our  simulations. Note that the total mass of the disc relates to the solid mass of the disc through 
the gas-dust ratio. The rectangle surrounded by a black box schematically shows the parameter space explored. The grey rectangle depicts the region where giant planets do not form at all.  The region of the parameter space where protoplanets reached the cross-over mass are shown in yellow for protoplanets accreting 1 km planetesimals, and in red for protoplanets accreting 0.1 km planetesimals. For bigger accreted planetesimal (100 and 10 km) no giant planets form. }
    \label{Fig:disc1}
    \end{figure}

  The parameter space explored for the initial mass and lifetime of protoplanetary discs is schematically shown in Fig. \ref{Fig:disc1}. The rectangle surrounded by a black box shows  the whole set of initial conditions studied. Note however that the distribution of these parameters is likely not uniform, and not all combinations have the same probability of occurrence: long lived discs, as well as very low and very massive discs are unlikely. We found that giant planets\footnote{By giant planets we mean protoplanets with masses larger than the cross-over mass and which survive from disc-planet angular momentum exchange and migration.}  do not form for the initial conditions corresponding to the grey region in Fig. \ref{Fig:disc1}. In fact, for a population of 100 km and 10 km planetesimals, giant planets do not form under any of the initial conditions considered in this work.  In the case of 100 km population of planetesimals this is due to the fact that  the accretion time to form a massive solid core is longer than the discs' lifetimes. For 10 km planetesimals,  some giant planets would form if migration were not at work (see Sec. \ref{discussion}). However, according to disc-planet interaction models, planets do migrate  and  the migration timescale 
turns out to be much shorter than the accretion timescale. Planets are lost in the central star before they are able to accrete enough solids to trigger the runaway accretion of gas (as seen in the previous section, 
this would allow them to grow faster and switch from type I to type II migration, therefore preventing their loss in the central star).
 

However, the situation reverts when the accreted planetesimals are smaller. Fig. \ref{Fig:disc1} shows that massive discs favour the formation of giant planets. The coloured regions depict the characteristics of discs where, for a particular initial position, the embryo succeeded in growing to a giant planet. This does not mean that a giant planet will form in any location of the disc, but that formation is possible at certain locations.  The yellow region shows the parameter space of  giant planets that succeeded to survive in the disc accreting 1 km planetesimals. Clearly, there is a dependence on the discs lifetime: less massive but longer lived discs favour the formation of giant planets. In red is plotted the situation for the case of $r_m = 0.1$ km accreted planetesimals. As smaller planetesimals are accreted faster and more efficiently,  the disc parameter space that leads to the formation of giant planets is bigger. Interestingly, in this case there is no dependence on the lifetime of the disc. When the amount of solids present is large enough, accretion to form a core with a critical 
mass proceeds so fast that it is always shorter than the lifetime of the discs considered here.  If the planetesimal disc is massive, accretion can be fast enough to enable protoplanets to reach the cross-over mass.  
Massive protoplanets switch from fast type I migration to a much slower type II, therefore decelerating and eventually braking before reaching the central star, preventing this way the loss of planets.  

Very small mass planets are the most abundant outcome in all simulations (regardless the size of the accreted planetesimals) and their final location could be  anywhere in the disc. Planets more massive than 
10 M$_\oplus$, in general, start their formation beyond the ice line (where solids are more abundant and feeding zones are larger), and due to migration they reach the inner regions of the disc. The final location of 
these planets extends between the inner edge of the disc and 10 AU.  Planets in the range of $10^2$ to $10^3$ M$_\oplus$ are the less abundant ones, and their final locations are very different from their initial 
emplacements. These planets undergo a lot of migration, and those that remain are the ones that were able to accrete gas fast enough to 
enter the runaway accretion of gas which prevented their  loss in the central star. Planets in the mass range 10 -  $10^2$ M$_\oplus$ are those that  undergo the largest net displacement from their original location. Most 
of the surviving planets in this mass range  have masses lower than 20 M$_\oplus$.  These planets, that were not able to reach their cross-over masses while they were migrating towards the central star are preserved 
because the disc dissipated before they could fall into the star.  In the case of $r_m= 0.1$ km, $\sim 18\%$ of the simulations lead to lost planets, 85\% of which were not able to reach their cross-over mass. This confirms 
our analysis of the previous section. Planets that are massive enough to undergo rapid inward type I migration ($\gtrsim 10$ M$_\oplus$) but whose growth rate is still dominated by the accretion of solids are 
likely to be lost in the star. 
  
In the previous sections we have noticed that results of giant planet formation depend upon the dynamical model adopted to describe planetesimals' dynamics. The equilibrium approach has the advantage of being 
numerically not very time consuming. However, it can lead to very different results when compared to the case of explicitly solving the differential equations for $e$ and $i$. When solving the equations, we also have 
the problem of setting the initial conditions, which are unknown. To test the importance of these assumptions, we performed $10\, 000$ simulations under three different conditions. Each calculation 
differs from the other only in the treatment of the planetesimals $e$ and $i$. This means that for a given set of initial conditions, we calculate the formation of the planet three times: one assuming the full 
equilibrium situation for planetesimals, another solving the explicit differential equations using as initial conditions for $e$ and $i$ the equilibrium value when the stirring timescale of the embryo equals the nebular 
gas drag timescale (we call this scenario ``hot disc'') and finally also solving the differential equations but using as initial conditions the equilibrium value between mutual planetesimal stirring and gas drag damping
(we call this scenario ``cold disc''). 

\begin{figure}
\centering
 \includegraphics[width= 0.4\textwidth, angle = -90]{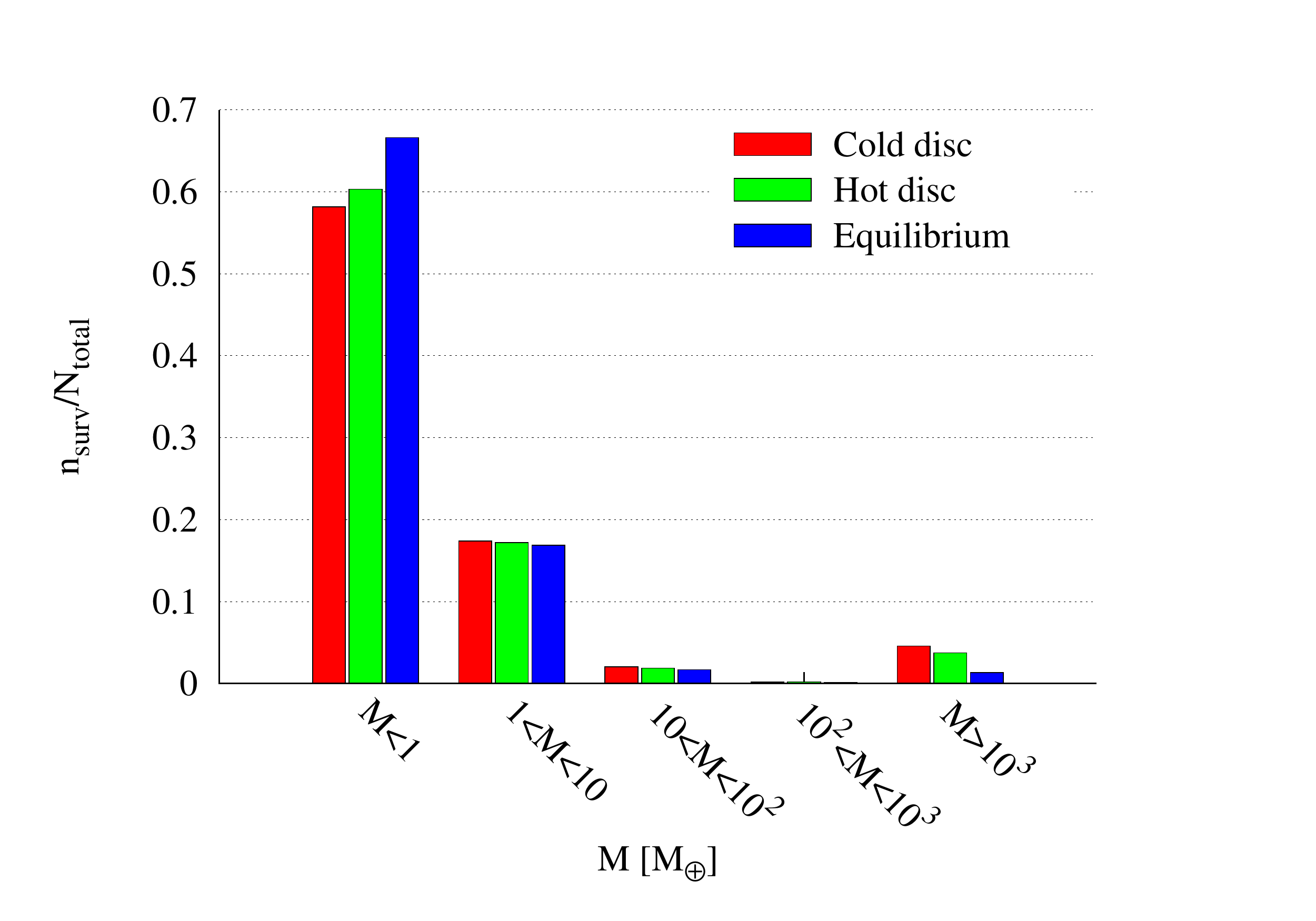} 

   \caption{Fraction of surviving planets with final mass in a certain mass interval. $\mathrm {N_{total}}$ is the sum of all the surviving planets ($\mathrm{n_{surv}}$) plus the planets lost in the central star. Final masses of the formed planets are separated in five mass bins (M<1 M$_\oplus$, 1<M$[M_\oplus]$<10, 10<M$[M_\oplus]$<100, 100<M$[M_\oplus]$<1000, M>1000 M$_\oplus$). The solid component of the planets grow by the accretion of  0.1 km planetesimals. Three different approaches in the treatment of the evolution of the eccentricities and inclinations of planetesimals are considered: equilibrium, and explicit calculation of the differential equations  for a ``hot'' and ``cold'' initial conditions.}
    \label{Fig:histogram1}
    \end{figure}

Fig. \ref{Fig:histogram1} shows the fraction of surviving planets with respect to the total amount of planets formed (surviving planets plus planets lost in the central star). Accreted planetesimals are 0.1 km size. To make this plot we 
classified the planets according to their final mass in five mass bins. Clearly, in all the cases, the most abundant planets are the low mass ones (< 1 M$_\oplus$). The equilibrium scenario being the one corresponding to the slowest accretion rate, is the one with more planets  in the lowest mass bin (< 1 M$_\oplus$). On the other extreme, in the case of the cold initial disc, accretion of solids at the beginning is more efficient, embryos grow bigger in a shorter time which, in turn, gives them more chances to continue accreting. That is why in the case of an initial cold disc planets are more massive. This is evident in the histogram: for a cold initial planetesimal disc,  the fraction of planets in each mass bin is the highest (except, of course, in lowest mass bin).  The amount of planets in the interval of 10 to $10^2$  M$_\oplus$  represents around  2\% of the surviving planets. Most of them are in the mass range of 10 to 20 M$_\oplus$, and just a few in the range of 20 to 50 M$_\oplus$. There are no planets in the range of 50 to $10^2$  M$_\oplus$, evidencing the dramatic effect of type I migration on these intermediate mass planets. In the mass interval $10^2$ - $10^3$ M$_\oplus$ it can be noticed a profound decrease in the number of planets, independently of the approach used for the planetesimal dynamics. This  can be understood as follows: protoplanets with masses greater than $10^2$ M$_\oplus$ are usually in the runaway phase of gas accretion. This means that hundreds of Earth masses can be accreted in a very short time. Therefore planets in the runaway phase easily grow more massive than Jupiter. Only if the disc dissipates during this process of accretion, final planetary masses can be between Saturn and a few Jupiter masses. To a less extent, planet migration has also some consequences in depleting this bin: although the mass bin in which planet migration is most effective in eliminating planets is the $10$ - $10^2$ M$_\oplus$ range, it is still very important in this mass regime, eliminating around 25\% of these planets (especially those closer to the $10^2$ M$_\oplus$  edge of the mass interval).

\section{Discussion}
\label{discussion} 
 
The calculations presented in this paper focus on the formation of planets and how this process is highly affected by the accretion rate of solids. The accretion rate of solids introduced intends to be realistic while computationally tractable. Some other aspects of the model are therefore rather simplified here;  for example
we consider the formation of a single planet. The application of these calculations to the formation of planetary systems will be presented elsewhere (Alibert et al. 2012). We introduced a 
semi-analytical description for the eccentricities and inclinations of planetesimals. In this work we calculated explicitly planetesimals'  eccentricities and inclinations, taking into account the
stirring of the growing planets, the gas drag from the nebula and the mutual stirring of the planetesimals themselves. The stirring produced by the growing planet excites planetesimals 
and makes their accretion more difficult as it grows. On the other hand, gas drag counteracts the stirring effect, an effect more important for smaller planetesimals. We have considered 
three approaches to determine the rms eccentricity and inclination of planetesimals:
 \begin{itemize}
  \item ``analytical equilibrium'' calculation, as described by Eqs. (\ref{eq:e_eq}) - (\ref{eq:i_eq}), 
  \item out of equilibrium, by solving the time evolution of the rms eccentricity and inclination of planetesimals (Eqs. (\ref{eq:diff_eqe}) - (\ref{eq:diff_eqi})), 
  starting from a ``cold'' planetesimal disc (their excitation state being the results of planetesimal-planetesimal interaction and gas drag only), and 
  \item out of equilibrium, starting from a ``hot'' disc, where planetesimals are already excited by the planetary embryo.
  \end{itemize}
  We have shown that these three approaches lead to different accretion rates, the difference depending on the planetesimal size, and being more important
  as a result of the migration and disc evolution feedbacks. For large planetesimals, the three approaches lead to similar results, but the accretion rate of solids
  under this hypothesis is very small, preventing the formation of massive planets. On the other hand, for low mass planetesimals, the excitation state of planetesimals
  as derived from the  ``analytical equilibrium'' calculations, and the excitation state of planetesimals following the second or third approach, even after a long time (when
  the equilibrium solution of Eqs. (\ref{eq:diff_eqe}) - (\ref{eq:diff_eqi}) is reached), are different, resulting in different accretion rate of solids. In particular, the ratio
  of the eccentricity to the inclination can deviate substantially from the 1/2 ratio assumed in the first approach.
  
  As a consequence of the size dependence of gas drag, small planetesimals are easier to accrete, leading to a faster  formation of planets. We have shown that 
  in the framework of the model hypothesis outlined in Sect. \ref{sec:formationmodel},  the formation of planets ranging from a fraction of an Earth mass to  several Jupiter masses  
  can only be accomplished under the assumption of a population of small planetesimals or, more generally, if planetesimals are, by any process, maintained in a  ``cold'' dynamical state. 
  Similar conclusions were already identified in other works (e.g. Fortier et al. 2007, 2009, Benvenuto et al. 2009), where it was shown that, in order to achieve the formation of the giant 
  planets of the solar system in less than 10 Myr most of the accreted planetesimals have to be small.  Those models, however, assumed in situ formation, and did not consider 
  a consistent calculation of the planet's final mass (the computations were stopped when the masses of the giant planets of the solar system were reached). Although 
  the approach for the calculation of planet formation (regarding the accretion of gas and solids) is similar,  our work  accounts for the migration of the planets and their 
  final masses are determined by the coupled evolution of the disc and the planet. 
 
\begin{figure*}
\centering
\includegraphics[width= \textwidth]{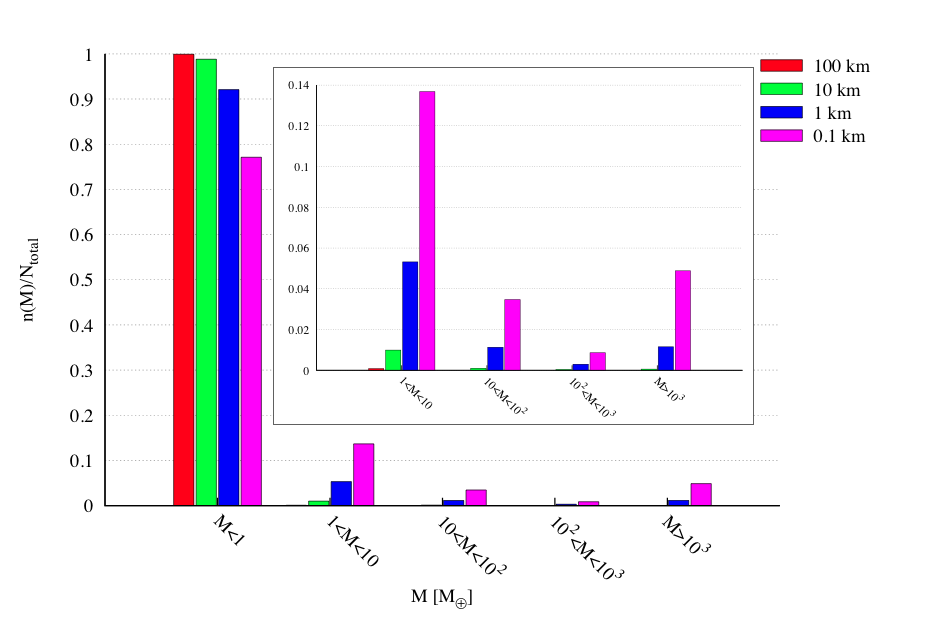} 

\caption{In situ formation. Fraction of planets with final mass in a certain mass interval.  $\mathrm{N_{total}}$ represents the total amount of planets formed, independently of their mass. Planets  are separated in five mass bins (M<1 M$_\oplus$, 1<M$[M_\oplus]$<10, 10<M$[M_\oplus]$<100, 100<M$[M_\oplus]$<1000, M>1000 M$_\oplus$). Accreted planetesimals have a radius of 100 km (red), 10 km (green), 1 km (blue) and 0.1 km (magenta). Most of the final masses are in the first mass bin. The inset caption represents the magnified histogram for the other mass bins.  }
  \label{Fig:histogram2}
\end{figure*}

 In order to compare our results with the afore-mentioned works, we performed simulations where migration is switched off. We calculate the in situ formation of $10\,000$ planets that accrete 
 planetesimals of 100, 10, 1 and 0.1 km in radius.  Indeed, as can be seen from Fig.  \ref{Fig:histogram2},  small planets are the most abundant ones. Larger planets form when there are small 
 planetesimals to accrete (smaller than $r_m = 10$ km). Planets in the mass range $10\, - \,10^2\, \mathrm{M}_\oplus$ (which growth is not yet dominated by the accretion of gas)  are produced in about the same 
 fraction as giant planets in the mass range  $> 10^3\,  \mathrm{M}_\oplus$, which growth is dominated by the a runaway accretion of gas. The runaway gas accretion leads to the difficulty of forming planets
 in the mass range  of  $10^2\, - \,10^3\, \mathrm{M}_\oplus$: only a fine tuned timing effect, the gas disc disappearing when the planet is in this mass range, can lead to planets in the Saturn-Jupiter domain.

 If we focus on the mass bin 10 to $10^2\, \mathrm{M}_\oplus$,  one difference between the migration and in situ scenario is that, while in the former there are no planets in the mass range between 50 and 100  $\mathrm{M}_\oplus$ (because they are all engulfed by the central star), in the in situ case planets are distributed in the whole mass spectrum of the bin (however, most of them are in the lowest mass range [10 to 20  $\mathrm{M}_\oplus$]). It is also interesting to compare for these two scenarios the mass percentage of solids for planets between $5\, - \,10^2 \, \mathrm{M}_\oplus$. As it can be seen in table \ref{table:core}, planets that migrate have a more massive core than planets formed in situ.  In general, when planets migrate they have access to regions of the disc that have not been depleted of planetesimals. Therefore, the solids surface density is higher and so is the accretion rate, which results in the formation of a more massive core.

\begin{table}
\caption{\label{table:core} For planets whose total mass is between 5 and $10^2$ $\mathrm{M}_\oplus$, this table shows the percentage of the total mass contained in the solid core. A comparison between a migration and non migration scenario is also made. In the migration case there are no planets with masses greater than 50 $\mathrm{M}_\oplus$.}
\centering
\begin{tabular}{p{2cm}  p{2cm}  p{1.8cm} }
\hline
Mass Interval & \multicolumn{2}{p{4 cm}}{ $\; \; \; \;\; \;\mathrm{M_{core}}/\mathrm{M_{total}} \times 100$} \\
$\; \; \; \; \; \;[\mathrm{M}_\oplus]$ & With Migration  & In situ\\
\hline
$\; \; \;\;\;5 - 10$  & $\; \; \;\;98$  &  $\; \;94$  \\
$\; \; \;10 - 20$   & $\; \; \;\;95$  &  $\; \;86$  \\
$\; \; \;20 - 30$   & $\; \; \;\;88$  &  $\; \;75$  \\
$\; \; \;30 - 40$   & $\; \; \;\;78$  &  $\; \;63$  \\
$\; \; \;40 - 50$   & $\; \; \;\;72$  &  $\; \;56$  \\
$\; \; \;50 - 60$   & $\; \; \;\; \; -    $  &  $\; \;56$  \\
$\; \; \;60 - 70$   & $\; \; \;\; \; -    $  &  $\; \;43$  \\
$\; \; \;70 - 80$   & $\; \; \;\;\;  -    $  &  $\; \;37$  \\
$\; \; \;80 - 90$   & $ \; \; \;\;\; -    $  &  $\; \;34$  \\
$\; \; \;90 - 100$ & $ \; \; \;\;\; -    $  &  $\; \;39$  \\
\hline
\end{tabular}
\end{table}

  In this paper we have considered  that the opacity of the envelope corresponds to the full interstellar medium opacity. However, several works (e.g. Pollack et al. 1996, Podolak 2003, Hubickyj et al. 2005, Movshovitz \& Podolak 2008) suggest that opacity in the planet's envelope should be much smaller, leading to a faster formation. As a complementary result, we computed one in situ population and one population including migration (\mbox{10 000 planets each}) reducing the opacity to 2\% of that of the interstellar medium. We found that with a reduced opacity the loss of planets decreases, increasing in every mass bin the number of planets that survive in the protoplanetary disc. Nevertheless, the shape of the mass distribution is very similar to that shown in  Fig.  \ref{Fig:histogram1}:  the fraction of surviving planets decreases with the mass of the planet, with a minimum in the interval $10^2-10^3\; \mathrm{M}_\oplus$ (although with the reduced opacity the fraction of planets in this bin is four times larger than with the full opacity), and raising again in the last mass interval. In table \ref{table:opa} we show the solids mass fraction for planets with masses between 5 to 100 $\mathrm{M}_\oplus$, considering both in situ and migration scenario. In both cases, the fraction of solids is smaller than in the corresponding case calculated with full opacity.  

\begin{table}
\caption{\label{table:opa} The same as in \ref{table:core} but considering that the planet's envelope opacity is only 2\% of the interstellar medium opacity.}
\centering
\begin{tabular}{p{2cm}  p{2cm}  p{1.8cm} }
\hline
Mass Interval & \multicolumn{2}{p{4 cm}}{ $\; \; \; \;\; \;\mathrm{M_{core}}/\mathrm{M_{total}} \times 100$} \\
$\; \; \; \; \; \;[\mathrm{M}_\oplus]$ & With Migration  & In situ\\
\hline
$\; \; \;\;\;5 - 10$  & $\; \; \;\;96$  &  $\; \;80$  \\
$\; \; \;10 - 20$   & $\; \; \;\;90$  &  $\; \;72$  \\
$\; \; \;20 - 30$   & $\; \; \;\;79$  &  $\; \;62$  \\
$\; \; \;30 - 40$   & $\; \; \;\;71$  &  $\; \;58$  \\
$\; \; \;40 - 50$   & $\; \; \;\;68$  &  $\; \;46$  \\
$\; \; \;50 - 60$   & $\; \; \;\; \; 50$  &  $\; \;44$  \\
$\; \; \;60 - 70$   & $\; \; \;\; \; 51$  &  $\; \;34$  \\
$\; \; \;70 - 80$   & $\; \; \;\;\;  48 $  &  $\; \;29$  \\
$\; \; \;80 - 90$   & $ \; \; \;\;\; -    $  &  $\; \;32$  \\
$\; \; \;90 - 100$ & $ \; \; \;\;\; 32   $  &  $\; \;20$  \\
\hline
\end{tabular}
\end{table}

Using the full model (including migration and disc evolution), giant planet formation by accretion of 100 km planetesimals is quite unlikely, if not impossible. In our simulations,  
to actually form giant planets we had to reduce the planetesimal size to the 0.1 km - 1 km radius range. Such a conclusion raises the question of the most likely planetesimal size
during the epoch of planet formation. Recent studies on planetesimal formation however give different conclusions. On one hand, models that explain the formation of planetesimals by
direct collapse in vortices in turbulent regions (e.g. Johansen et al. 2007) predict a fast formation of very big planetesimals ($r_m > 100$ km). We note however that this formation
process may not be totally efficient and only a small fraction of solids initially present in protoplanetary discs is likely to end up in such big planetesimals. The conclusions of Johansen
et al. (2007) are consistent with the results of Morbidelli et al. (2009) on the initial function of planetesimals. On the other hand, a recent study of Windmark et al. (2012) shows 
that direct growth of planetesimals via dust collisions can lead to the growth of 0.1 km  planetesimals.  In addition, initially small planetesimals show better matches to the observed 
size distribution of objects in the asteroid belt and among the TNOs:  Weidenschilling (2011) shows that the size distribution currently observed in the asteroid belt in the range of 
10 to 100 km can be better explained by an initial population of 0.1 km planetesimals. Kenyon and Bromley (2012) conclude, by combining observations of the hot and cold 
populations of TNOs with time constraints on their formation process, that TNOs form from a massive disc mainly composed of 1 km planetesimals. More investigations on the 
formation of planetesimals, and planetary embryos, are definitely required in order to test the viability of planetary core formation by accretion of low mass planetesimals. We 
note finally that what is important in the work we have presented now is \textit{not} the initial mass function of planetesimals, but their mass function at the time of planet formation. 
The two quantities are likely to differ due to planetesimal-planetesimal collisions, and the resulting mass growth and/or fragmentation.

\section{Conclusions }
\label{conclusions} 
 
In this paper we presented calculations of planetary formation considering the formation of a single planet at each time, and starting with embryos of $0.01\, \mathrm{M}_\oplus$. 
Our simulations consist in the calculation of the formation of a planet, including its growth in mass by accretion of solids and gas, its migration in the disc, and the evolution of the disc until the gas component
 of the disc is dispersed. When the nebular gas is gone, simulations are stopped. Therefore, the subsequent growth of the planets, by accretion of residual planetesimals or collisions among embryos
 if we were considering planetary systems, is not considered. During their formation, the growth of the planets is calculated self-consistently coupling the accretion of solids and gas. The accretion 
 of solids is computed assuming the particle-in-a-box approximation, and computing the excitation state of planetesimals, which in turn regulates to a great extent the accretion rate of the planet. 
 The accretion of gas is computed solving the differential equations that govern the evolution of the structure of the planet. Finally, protoplanets grow in an evolving protoplanetary disc, which 
 density, temperature and pressure is calculated at every time  step.

The  combination of oligarchic growth (for the solid component  of the planet) with the migration of the planet has severe consequences for protoplanets 
that are able to grow up to a  few tens Earth masses: these planets tend to collide with the central star (or at least to migrate to the innermost location of the protoplanetary disc). 
Indeed, planets that are between 10 and 100 Earth  masses are usually undergoing very rapid inward type I migration, but are not massive enough to switch to a slower, type II  
migration. In our simulations, the only surviving planets in the range of 10 to 20 Earth masses correspond to cases where the gas component of the  disc dissipates during 
their growth,  preventing them to fall into the star. On the opposite, if the solid core grows fast enough, it enables the accretion of large amounts of gas, when the critical mass 
is reached. At this point,  the runaway of gas ensures an extremely quick growth in the mass of the planet, and the planet migration rate decreases.  

In the model we have presented in this paper, we have assumed a uniform population of small planetesimals which size remains unchanged during the whole formation of the planet.
Most probably the initial population of planetesimals in protoplanetary discs is not uniform in size, but follows a size distribution. 
We have  shown, however,  that without small planetesimals 
giant planet formation is difficult to explain, at least in the way we understand it now. However, even with an initial population of small planetesimals, the collisions among themselves are 
likely to be disruptive as soon as their random velocities start to be excited by a planetary embryo.Therefore, it is also unlikely that an initial population of only small planetesimals can be 
used to explain the formation of giant planets. Moreover,  even assuming this to be true, only a few amount of planets in the range of several tens Earth masses to 
a few Jupiter masses can be formed. 
   
In addition, small mass planetesimals are subject to large radial drift, as a result of gas drag. Planetesimal drift can have positive or negative consequences in the formation of 
planetary systems, as has been shown by Guilera et al. (2010, 2011). Similarly, fragmentation and coagulation can hasten or delay planet formation as a whole: fragments of 
smaller mass are easier to accrete but, if they also can leave a planet's feeding zone very quickly as a result of gas drag. Finally, in a planetary system, fragments that are 
not accreted by the embryo that generated them can be accreted by another embryo located in an innermost region. It is not clear what the possible outcomes of putting together 
planetesimals drifting, fragmentation and many embryos forming in an evolving disc could be. This however represents a very important step in the understanding of the first 
stage of planet formation.
       
Because most of the accretion of solids should, at some point, be dominated by small planetesimals or fragments, our calculations can be understood as a description of that stage. 
An interesting scenario to analyse, in particular if planetesimals are born massive (Johansen et al. 2007, Morbidelli et al. 2009) would be the following. An initial population dominated 
by $\sim 100$ km planetesimals would prevent a fast growth of the embryos at the beginning, a time during which the planetary embryos would suffer only of a little migration and 
the protoplanetary disc would evolve, reducing its gas surface density. Fragments (smaller planetesimals), resulting from collisions between big planetesimals,  would start 
to be generated later (the timescale of fragmentation of 100 km planetesimals affected by the stirring of Moon to Mars mass embryos is of the order $10^6$ years), therefore accelerating 
the formation of the embryo in a later stage of the disc evolution. The collisional cascade would probably still produce small fragments  fast enough to help the growth of an embryo,
even if they leave the feeding zone very fast. Therefore, protoplanets could grow by the accretion of fragments, not necessarily generated by themselves, but generated by another distant 
embryo. Putting together these different processes  will give us a better insight on the formation process and would help us to constrain, from 
planetary formation models, possible initial size distributions for planetesimals.  

It is also important to mention that we have not considered the possibility of planetesimal-driven migration. Although in general planetesimal-driven migration acts on a longer timescale that Type I migration, recently Ormel et al. (2012) find that planetesimal-driven migration can have a mild effect on mid-sized planets in massive planetesimal discs, competing with Type I migration.

The main conclusions of our work are that formation of giant planets in the framework of the sequential accretion model needs the presence of unexcited planetesimals. One obvious way 
to de-excite planetesimals is through gas drag, but this requires this latter to be efficient, which in turn translates to low mass planetesimals. These planetesimals can be primordial 
or fragments of originally bigger ones. But, at some point,  small boulders are needed to build protoplanetary massive cores before the dissipation of the disc. The combination of 
migration and oligarchic growth, on the other hand, prevents the formation of intermediate mass planets. However, this result can change when considering the formation of planets 
in planetary systems where their gravitational interactions are taken into account. Captures in resonances can prevent planets from colliding with the central star, preserving them 
in planetary systems. Exploring this effects will be the subject of future works.

\begin{acknowledgements}
     This work was supported by the European Research Council under grant 239605, the Swiss National Science Foundation. We thank C. Mordasini  for the fruitful discussions. We also wish to thank  the referee, J.  Lissauer,  for constructive criticisms which enabled us to improve the original version of this article.
\end{acknowledgements}

\end{document}

%% file: def.tex
\def\mearth{M_{\oplus}}

\def\tdisc{T_{\rm disc}}
\def\tsurf{T_{\rm surf}}

\def\f1{f_{\rm I}}

\def\beq{\begin{equation}}
\def\eeq{\end{equation}}

\def\cs{C_s}

\def\mdisc{M_{\rm disc}}
\def\ac{a_{\rm C}}

\def\pcore{P_{\rm core}}
\def\tsurfpla{T_{\rm surf, planet}}
\def\psurfpla{P_{\rm surf, planet}}
\def\tdisc{T_{\rm disc}}
\def\pdisc{P_{\rm disc}}
\def\hdisc{H_{\rm disc}}
\def\rhodisc{\rho_{\rm disc}}
\def\tauout{\tau}
\def\vff{v_{\rm ff}}
\def\mdotgasmax{\dot{M}_{\rm gas, max}}

\def\mtot{M_{\rm tot}}

\def\mearth{M_\oplus}

\def\rhill{R_{\rm Hill}}

\def\rplanet{R_{\rm planet}}

\def\mplanet{M_{\rm planet}}

\def\mcore{M_{\rm core}}
\def\rcore{R_{\rm core}}
\def\mearth{M_\oplus}

\def\mcore{M_{\rm core}}

\def\rhill{R_{\rm H}}

\def\simgr{\,\hbox{\hbox{$ > $}\kern -0.8em \lower 1.0ex\hbox{$\sim$}}\,}
\def\simle{\,\hbox{\hbox{$ < $}\kern -0.8em \lower 1.0ex\hbox{$\sim$}}\,}
\def\beq{\begin{equation}}
\def\eeq{\end{equation}}
\def\dpartial#1#2{{{\partial {#1}} \over {\partial {#2}}}}

\def\simgr{\,\hbox{\hbox{$ > $}\kern -0.8em \lower 1.0ex\hbox{$\sim$}}\,}
\def\simle{\,\hbox{\hbox{$ < $}\kern -0.8em \lower 1.0ex\hbox{$\sim$}}\,}
\def\beq{\begin{equation}}
\def\eeq{\end{equation}}

\def\aj{AJ}                   
\def\apj{ApJ}                 
\def\apjl{ApJ}                
\def\apjs{ApJS}               
\def\ao{Appl.Optics}          
\def\aap{A\&A}                
\def\mnras{MNRAS}             

\def\({\left(}
\def\){\right)}
\def\<{\left<}
\def\>{\right>}

\def\({\left(} 
\def\){\right)} 
\def\<{\left<} 
\def\>{\right>} 

\def\bc{\begin{changebar}}
\def\bce{\begin{center}}
\def\beq{\begin{equation}} 

\def\bi{\begin{itemize}}

\def\btab{\begin{tabular}{p{1.7cm}p{12cm}p{1.5cm}}}
\def\bt2{\begin{tabular}{p{1 cm}p{4.5cm}p{10cm}}}
\def\cs{c_\mathrm{s}}

\def\dpartial#1#2{{{\partial {#1}} \over {\partial {#2}}}} 

\def\ec{\end{changebar}}
\def\ece{\end{center}}
\def\eeq{\end{equation}} 
\def\ei{\end{itemize}}

\def\etab{\end{tabular}\\}

\def\mH2{m_\mathrm{H_2}}

\def\dmax0{\rho_\mathrm{max}}
\def\dmaxS0{\Sigma_\mathrm{max}}

\def\rH2{r_\mathrm{H_2}}

\def\r0max{r_\mathrm{0max}}

\def\s0{\sigma_\mathrm{0}}

\def\xp0{x_{\rm{M0}}}

\def\z0max{z_\mathrm{0max}}

\def\aj{AJ}                  
\def\apj{{\it ApJ}}                 
\def\apjl{ApJ}                
\def\apjs{ApJS}               
\def\ao{Appl.Optics}          
\def\aap{{\it A\&A}}                
 
\def\mnras{MNRAS}             

\def\nat{Nature}

%% file: fortier_article_version.bbl
\begin{thebibliography}{}

\bibitem[1976]{adachi} Adachi, I., Hayashi, C., \& Nakazawa, K.\ 1976, Progress of Theoretical Physics, 56, 1756 

\bibitem[2005a]{alibertetal2005a}
  Alibert, Y., Mordasini, C., Benz, W., \& Winisdoerffer, C. 2005a, \aap,  434, 343

\bibitem[2005b]{alibertetal2005b}
  Alibert, Y., Mousis, O., Mordasini, C. \& Benz, W. 2005b, \apj,  626, L57
  
\bibitem[2010]{Andrews} Andrews, S.~M., Wilner, D.~J., Hughes, A.~M., Qi, C., \& Dullemond, C.~P.\ 2010, \apj, 723, 1241 

\bibitem[1994]{Bell} Bell, K.~R., \& Lin, D.~N.~C.\ 1994, \apj, 427, 987 

\bibitem[2009]{bfb} Benvenuto, O.~G., Fortier, A., \& Brunini, A.\ 2009, Icarus, 204, 752 

\bibitem[2006]{chambers} Chambers, J.\ 2006, Icarus, 180, 496 

\bibitem[2006]{Crida} Crida, A., Morbidelli, A., \& Masset, F.\ 2006, \icarus, 181, 587 

\bibitem[2007]{fbb} Fortier, A., Benvenuto, O.~G., \& Brunini, A.\ 2007, \aap, 473, 311 

\bibitem[2009]{fbb2} Fortier, A., Benvenuto, O.~G., \& Brunini, A.\ 2009, \aap, 500, 1249

\bibitem[2010]{Guilera2010} Guilera, O.~M., Brunini, A., \& Benvenuto, O.~G.\ 2010, \aap, 521, A50 

\bibitem[2011]{Guilera} Guilera, O.~M., Fortier, A., Brunini, A., \& Benvenuto, O.~G.\ 2011, \aap, 532, A142 

\bibitem[2005]{Hubickyj} Hubickyj, O., Bodenheimer, P., \& Lissauer, J.~J.\ 2005, \icarus, 179, 415 

\bibitem[1993]{ida} Ida, S., \& Makino, J.\ 1993, \icarus, 106, 210 

\bibitem[2004a]{idalin2004a} Ida, S. \& Lin, D.N.C. 2004a, \apj  604, 388

\bibitem[Inaba \& Ikoma(2003)]{2003A&A...410..711I} Inaba, S., \& Ikoma, M.\ 2003, \aap, 410, 711 

\bibitem[2001]{Inaba} Inaba, S., Tanaka, H., Nakazawa, K., Wetherill, G.~W., \& Kokubo, E.\ 2001, \icarus, 149, 235 

\bibitem[2007]{johansen_2007} Johansen, A., Oishi, J.~S., Mac Low, M.-M., Klahr, H., Henning, T., \& Youdin, A.\ 2007, \nat, 448, 1022 

\bibitem[Kenyon \& Bromley(2012)]{2012AJ....143...63K} Kenyon, S.~J., \& Bromley, B.~C.\ 2012, \aj, 143, 63 

\bibitem[2006]{kley} Kley, W., \& Dirksen, G. \ 2006, \aap, 447, 369




\bibitem[2009]{Lissauer} Lissauer, J.~J., Hubickyj, O., D'Angelo, G., \& Bodenheimer, P.\ 2009, \icarus, 199, 338 


\bibitem[2009]{Mamajek} Mamajek, E.~E.\ 2009, American Institute of Physics Conference Series, 1158, 3 

\bibitem[1995]{Mayor} Mayor, M., \& Queloz, D.\ 1995, \nat, 378, 355 

\bibitem[2011]{Min} Min, M., Dullemond, C.~P., Kama, M., \& Dominik, C.\ 2011, \icarus, 212, 416 

\bibitem[1980]{Mizuno} Mizuno, H.\ 1980, Progress of  Theoretical Physics, 64, 544 


\bibitem[Morbidelli et al.(2009)]{2009Icar..204..558M} Morbidelli, A., Bottke, W.~F., Nesvorn{\'y}, D., \& Levison, H.~F.\ 2009, \icarus, 204, 558 

\bibitem[2009]{mordasini2009a} Mordasini, C., Alibert, Y. \& Benz, W. 2009, \aap,   510, 1139

\bibitem[2010]{Mordasini} Mordasini, C., Klahr, H., Alibert, Y., Benz, W., \& Dittkrist, K.-M.\ 2010, arXiv:1012.5281
 
\bibitem[2012]{Mordasini2012a} Mordasini, C., Alibert, Y., Benz, W., Klahr, H., \& Henning, T.\ 2012, \aap, 541, A97 

\bibitem[2010]{Mordasini2012b} Mordasini, C., Alibert, Y., Klahr, H., \& Henning, T. \ 2012b, aXiv: 1206.6103

\bibitem[2008]{Movshovitz} Movshovitz, N., \& Podolak, M.\ 2008, \icarus, 194, 368 

\bibitem[Ohtsuki et al.(2002)]{2002Icar..155..436O} Ohtsuki, K., Stewart, G.~R., \& Ida, S.\ 2002, \icarus, 155, 436 

\bibitem[2010]{ormel} Ormel, C.~W., Dullemond, C.~P., \& Spaans, M.\ 2010, \apjl, 714, L103

\bibitem[2012]{ormel2} Ormel, C.~W.,  Ida, S., \& Tanaka, H. \ 2012, arXiv: 1207.7104

\bibitem[2010]{Paardekooper} Paardekooper, S.-J., Baruteau, C., Crida, A., \& Kley, W.\ 2010, \mnras, 401, 1950 

\bibitem[2011]{Paardekooper2011} Paardekooper, S.-J., Baruteau, C., \& Kley, W.\ 2011, \mnras, 410, 293 


\bibitem[2003]{Podolak} Podolak, M.\ 2003, \icarus, 165, 428 

\bibitem[Pollack et al.(1996)]{1996Icar..124...62P} Pollack, J.~B., Hubickyj, O., Bodenheimer, P., et al.\ 1996, \icarus, 124, 62 

\bibitem[2004]{Rafikov} Rafikov, R.~R.\ 2004, \aj, 128, 1348 

\bibitem[1995]{SCVH} Saumon, D., Chabrier, G., \& van Horn, H.~M.\ 1995, \apjs, 99, 713 

\bibitem[1973]{SS}Shakura, N. I. \& Sunyaev, R. A. 1973, \aap, 24, 337

\bibitem[Stepinski \& Valageas(1996)]{1996A&A...309..301S} Stepinski, T.~F., \& Valageas, P.\ 1996, \aap, 309, 301 

\bibitem[2002]{tanaka} Tanaka, H., Takeuchi, T., \& Ward, W.~R.\ 2002, \apj, 565, 1257 

\bibitem[2003]{thommes} Thommes, E.~W., Duncan,  M.~J., \& Levison, H.~F.\ 2003, Icarus, 161, 431 

\bibitem[2010]{Valencia} Valencia, D., Ikoma, M., Guillot, T., \& Nettelmann, N.\ 2010, \aap, 516, A20 

\bibitem[Veras \& Armitage(2004)]{2004MNRAS.347..613V} Veras, D., \& Armitage, P.~J.\ 2004, \mnras, 347, 613 

\bibitem[1997]{1997ApJ...482L.211W} Ward, W.~R.\ 1997, \apjl, 482, L211 

\bibitem[Windmark et al.(2012)]{2012arXiv1201.4282W} Windmark, F., Birnstiel, T., G{\"u}ttler, C., et al.\ 2012, arXiv:1201.4282 

\bibitem[Weidenschilling(2011)]{2011Icar..214..671W} Weidenschilling, S.~J.\ 2011, \icarus, 214, 671 


\end{thebibliography}
